\documentclass[twocolumn]{aastex631}

\usepackage{multirow}
\usepackage{makecell}

\DeclareRobustCommand{\ion}[2]{%
\relax\ifmmode
\ifx\testbx\f@series
{\mathbf{#1\,\mathsc{#2}}}\else
{\mathrm{#1\,\mathsc{#2}}}\fi
\else\textup{#1\,{\mdseries\textsc{#2}}}%
\fi}

\newcommand\nh{\ifmmode{n_{\tiny \mbox{H}}}\else{$n_{\tiny \mbox{H}}$}\fi}
\newcommand\Nh{\ifmmode{N_{\tiny \mbox{H}}}\else{$N_{\tiny \mbox{H}}$}\fi}
\newcommand\ngr{\ifmmode{n_{\tiny \mbox{gr}}}\else{$n_{\tiny \mbox{gr}}$}\fi}
\newcommand\jvsp{\ifmmode{j_{\tiny \nu,sp}}\else{$j_{\tiny \nu, sp}$}\fi}
\newcommand\jvff{\ifmmode{j_{\tiny \nu,ff}}\else{$j_{\tiny \nu,ff}$}\fi}
\newcommand\jvbb{\ifmmode{j_{\tiny \nu,bb}}\else{$j_{\tiny \nu,bd}$}\fi}
\newcommand\amax{\if{a_{\tiny \mbox{max}}}\else{$a_{\tiny \mbox{max}}$}\fi}
\newcommand\amin{\if{a_{\tiny \mbox{min}}}\else{$a_{\tiny \mbox{min}}$}\fi}
\newcommand\cmvol{\ifmmode{\mbox{cm}^{-3}}\else{$\mbox{cm}^{-3}$}\fi}

\newcommand\hii{$\textrm{H}\scriptstyle\mathrm{II}$}
\accepted{August 23, 2024}

\shorttitle{Redshift Variation of Radio-IR Correlation}
\shortauthors{Yoon}

\begin{document}

\title{A simple model of the radio-infrared correlation depending on gas surface density and redshift}

\correspondingauthor{Ilsang Yoon}
\email{iyoon@nrao.edu}

\author[0000-0001-9163-0064]{Ilsang Yoon}
\affiliation{National Radio Astronomy Observatory, 520 Edgemont Road, Charlottesville, VA 22903, USA}

\begin{abstract}
We introduce a simple parametric model of the radio-infrared correlation (i.e., the ratio between the IR luminosity and the 1.4 GHz radio luminosity, $q_{\mbox{\tiny IR}}$) by considering the energy loss rate of high-energy cosmic ray (CR) electron governed by the radiative cooling (synchrotron, bremsstrahlung, inverse Compton scattering), ionization, and adiabatic expansion. Each process of CR electron energy loss is explicitly computed and compared to each other. We rewrite the energy loss rate of each process to be dependent on the gas surface density and redshift using the relevant scaling relations. By combining each energy loss rate, the fraction of the synchrotron energy loss rate is computed as a function of gas surface density and redshift, and used to extrapolate the well-established `local' radio-infrared correlation to the high-redshift universe. The locally established $q_{\mbox{\tiny IR}}$ is reformulated to be dependent upon the redshift and the gas surface density and applied for understanding the observed distribution of the radio-infrared correlation of high-redshift galaxies in \cite{delvecchio_etal_2021}. Our model predicts that $q_{\mbox{\tiny IR}}$ value is anti-correlated with gas surface density and the redshift dependency of $q_{\mbox{\tiny IR}}$ value changes by gas surface density of galaxies, which captures the observed trend of $q_{\mbox{\tiny IR}}$ values for stellar mass selected star forming galaxies with a minimal impact of radio-infrared selection bias.
\end{abstract}


\section{Introduction} \label{sec:intro}
The correlation between the non-thermal radio and IR luminosity in galaxies \citep[i.e., radio-IR correlation,][]{helou_etal_1985} is one of the tightest and best studied in astronomy and its nearly linear relationship held over 5 orders of magnitude implies a direct relationship between star formation and cosmic-ray production \citep{yun_etal_2001}: infrared emission comes from dust heated by massive OB stars, while non-thermal radio emission (i.e., synchrotron emission) arises from relativistic cosmic-ray electrons accelerated by shock waves produced when massive stars explode as supernovae \citep{delvecchio_etal_2021}. Nevertheless, these high energy cosmic-ray (CR) electrons are also subject to other cooling processes as they propagate throughout the galaxy, which are mainly caused by inverse Compton, bremsstrahlung, and ionization losses \citep{murphy_2009}.

Despite all these multiple processes governed by different spatial and time scales, the synchrotron-dominated radio emission at 1.4GHz has been widely used to investigate the star formation of galaxies in a dust unbiased way and also used as a star formation rate (SFR) estimator for high-$z$ galaxies from deep radio surveys. While the tight correlation between radio and IR luminosity is well established empirically in local universe \citep{yun_etal_2001,bell_2003}, the redshift variation of the radio-IR correlation is not well constrained: some studies suggest a systematic decrease of the radio-IR correlation with reshift \citep[e.g.,][]{seymour_etal_2009,basu_etal_2015b,calistro_rivera_etal_2017,delhaize_etal_2017} and others suggest a null correlation \citep[e.g.,][]{jarvis_etal_2010,sargent_etal_2010,bourne_etal_2011}. Systematic bias in the observational parameters, for example, galaxy type \citep{vollmer_etal_2022}, dust temperature \citep{smith_etal_2014} and star formation rate calibration \citep{molnar_etal_2021} may introduce a bias in sampling galaxies in redshift and affect the derived correlation. 

To take these kinds of selection bias into account, a recent study by \cite{delvecchio_etal_2021} samples the galaxies in a range of redshift ($0.1<z<4.5$) and stellar mass ($10^8<M_{*}/M_{\odot}<10^{12}$) independent of the radio and IR detection and shows that the radio-IR correlation varies as a function of galaxy stellar mass (and also star formation rate surface density due to the correlation between galaxy stellar mass and SFR) without a significant evolution with redshift for fixed stellar mass. Motivated by this observational study, \cite{schober_etal_2022} proposes a physical model of the radio-IR correlation including relevant physical processes (e.g., CR electron production and magnetic field generated by small-scale dynamo) for the main sequence galaxies to explain the dependency of radio-IR correlation on galaxy stellar mass seen in \cite{delvecchio_etal_2021}.  

Cosmic ray (CR) particles are high-energy electrons (e$_{\mbox{\tiny CR}}$) and nuclei (i.e., mostly protons, p$_{\mbox{\tiny CR}}$) injected from the energetic event (e.g., supernova explosion) that cools by the interaction with the surrounding medium \citep{longair_2011}. Such interaction falls into three categories: (1) losses due to electromagnetic interaction of the charged particles with the medium in which they propagate (i.e., ionization); (2) radiative processes; and (3) nuclear collision that produces secondary high energy CR particles. Due to its heavy mass, CR proton is not a major participant in the radiative processes and therefore primarily involved in the ionization process and the nuclear collision. In radiative processes of CR cooling, CR electron is a dominant species. Three major radiative processes that cool the CR electrons are synchrotron emission, inverse Compton scattering with ambient radiation field, bremsstrahlung emission. In addition, CR electrons can also cool by the ionization process and, in a certain case, by adiabatic expansion if the CR electrons stay in the CR electron cloud while it is expanding. To understand the relative importance of the synchrotron emission from CR electrons, we need to compare the energy loss rate of CR electron by each cooling processes.

A rigorous model of the radio-IR correlation requires a proper inclusion and tuning of multiple physical processes with different spatial and time scales \citep[e.g., ][]{lacki_etal_2010a,schober_etal_2022}, which are not fully understood by the observations. Therefore, a simple parametric form based on the global scaling relations of galaxy properties may be useful for understanding the general trend of the radio-IR correlation. 

In this study, we compare different CR electron cooling processes and develop a simple model of the radio-IR correlation parameterized by redshift and gas surface density which may introduce a galaxy selection bias in characterizing $q_{\mbox{\tiny IR}}$ distribution: galaxies with higher gas surface density are more likely to be detected in the radio-IR surveys than galaxies with lower gas surface density at a given redshift because of the correlation between gas surface density and star formation rate surface density \citep{ks_relation_2012} with caveats about a variation in slope and normalization amongst galaxies with different properties. 

Our model is based on simple scaling relations without having physical processes involving CR electron transport \citep[e.g.,][]{lacki_etal_2010a,lacki_etal_2010b} and production of secondary CR electrons, but tries to capture the observed trend of the radio-IR correlation, which may be useful for understanding the radio observations of galaxies in high-$z$ universe and for implementing a prescription of creating mock radio catalogs from galaxy formation simulations. In this paper, we assume a $\Lambda$CDM cosmology with $H_0=70\text{ km/s/Mpc}$, $\Omega_M=0.3$, and $\Omega_\Lambda=0.7$.

\section{Model}
\subsection{Radio-IR correlation}
The radio-IR correlation \citep[e.g.,][]{helou_etal_1985} between radio luminosity at 1.4 GHz and IR luminosity at 80$\mu$m (i.e., $3.75\times10^{12}$Hz) is defined as
\begin{equation}\label{eq:qir}
q_{\mbox{\tiny IR}}\equiv\mbox{log}\left(\frac{L_{\mbox{\tiny IR}}}{3.75\times10^{12}~\mbox{\small erg s}^{-1} \mbox{\small Hz}^{-1}}\right) - \mbox{log}\left(\frac{L_{\mbox{\tiny 1.4GHz}}}{\mbox{\small erg s}^{-1}\mbox{\small Hz}^{-1}}\right).
\end{equation} 

For optically thin thermal free-free radio emission, the ionizing photon rate is directly proportional to the thermal spectral luminosity, $L_\nu^T$ \citep[e.g.,][]{murphy_etal_2011,rubin_1968} as
\begin{eqnarray}\label{eq:photonrate}
\left[\frac{Q(H^0)}{\mbox{s}^{-1}}\right] & =  & 6.3\times10^{25} \epsilon \left(\frac{T_e}{10^4\mbox{K}}\right)^{-0.45} \left(\frac{\nu}{\mbox{GHz}}\right)^{0.1} \nonumber \\
& & \times \left(\frac{L_\nu^T}{\mbox{erg s}^{-1}\mbox{Hz}^{-1}}\right)
\end{eqnarray} where, by introducing a parameter $\epsilon \gtrsim 1.0$, we note that the ionizing photon rate can be underestimated by free-free emission if a significant fraction of ionizing photons is absorbed by dust \citep{murphy_etal_2011}. The ionizing photon rate can be related to the star formation rate (SFR) over the last 100 Myr with the assumed solar metallicity and continuous star formation \citep{murphy_etal_2011}
\begin{equation}\label{eq:sfrphot}
\left(\frac{\mbox{SFR}}{M_{\odot}\mbox{yr}^{-1}}\right) = 7.29\times10^{-54}\left[\frac{Q(H^0)}{\mbox{s}^{-1}}\right].
\end{equation}
The SFR over the last 100 Myr can be also related to the total IR luminosity \citep{calzetti_2013}
\begin{equation}\label{eq:sfrir}
\left(\frac{\mbox{SFR}}{M_{\odot}\mbox{yr}^{-1}}\right) = 2.8\times10^{-44} \left(\frac{L_{\mbox{\tiny IR}}}{\mbox{erg s}^{-1}}\right)
\end{equation}

Therefore, by combining Equation~\ref{eq:photonrate}, \ref{eq:sfrphot}, and \ref{eq:sfrir}, we obtain
\begin{eqnarray}\label{eq:sub}
\left(\frac{L_{\mbox{\tiny IR}}}{3.75\times10^{12}~\mbox{\small erg s}^{-1} \mbox{\small Hz}^{-1}}\right) & = & 4374 \epsilon \left(\frac{T_e}{10^4\mbox{K}}\right)^{-0.45} \nonumber\\
& & \times \left(\frac{\nu}{\mbox{GHz}}\right)^{0.1} \nonumber \\
& & \times \left(\frac{L_\nu^T}{\mbox{erg s}^{-1}\mbox{Hz}^{-1}}\right)
\end{eqnarray}

Using Equation~\ref{eq:sub} with $T_e=10^4$K for typical electron temperature in \hii\ region and $\epsilon=1.0$, we can express Equation~\ref{eq:qir} as 
\begin{equation}\label{eq:qir3}
q_{\mbox{\tiny IR}} = 3.64+0.1~\mbox{log}\left(\frac{\nu}{\mbox{GHz}}\right) + \mbox{log}\left(\frac{L_\nu^T}{L_{\mbox{\tiny 1.4GHz}}}\right)
\end{equation}

If we assume that the radio luminosity at 1.4GHz is a mixture of thermal free-free and non-thermal synchrotron emission with assumed thermal fraction $f_{th}$,   
\begin{equation}\label{eq:qir2}
    q_{\mbox{\tiny IR}}=3.64 + 0.1~\mbox{log}\left(\frac{\nu}{\mbox{GHz}}\right) + \mbox{log}\left(f_{th} \frac{L_\nu^T}{L_{\mbox{\tiny 1.4GHz}}^{T}}\right)
\end{equation} where $L_{\mbox{\tiny 1.4GHz}}^{T}$ is the spectral luminosity of thermal free-free emission at 1.4 GHz contributing to the total 1.4GHz radio luminosity $L_{\mbox{\tiny 1.4GHz}}$. Adopting the power-law free-free radio emission $F_\nu \sim \nu^{-0.1}$, $\nu=33$GHz for $L_\nu^T$ by assuming a negligible contribution from non-thermal synchrotron emission being dominant up to 10GHz \citep{klein_etal_2018}, 10\% thermal fraction at 1.4GHz ($f_{th}=0.1$), Equation~\ref{eq:qir2} results in $q_{\mbox{\tiny IR}}=2.65$ which is the average value ($q_{\mbox{\tiny IR}}=2.64\pm0.26$) obtained in the local universe \citep{bell_2003}. 

\subsection{Energy loss rate of CR electron and fraction of synchrotron cooling}
The $q_{\mbox{\tiny IR}}$ value may change with the redshift because of the relative increase of CMB energy density which increases the cooling of high-energy cosmic ray (CR) electrons by inverse Compton scattering (IC) resulting in a synchrotron `dimming'. However, the radio dimming cannot be described by a simple competition between magnetic field energy density and the CMB energy density; the CMB energy density must also compete with every other cooling processes \citep{lacki_etal_2010b}. To characterize and illustrate the significance of synchrotron emission compared to other CR electron cooling processes discussed in Section~\ref{sec:intro}, we compare the energy loss rate (erg s$^{-1}$) of CR electron via synchrotron, CMB IC, bremsstrahlung, ionization, and adiabatic expansion each of which is discussed in \cite{longair_2011}.  

The expression of the energy loss rate ($P=dE/dt$) of a CR electron moving with kinetic energy $E=(\gamma-1) m_e c^2$ via synchrotron, CMB IC, bremsstrahlung, ionization, and adiabatic expansion are shown in the following sub-sections.

\subsubsection{Synchrotron emission}
The energy loss rate of a single CR electron propagating through the ISM with a magnetic field, via synchrotron emission is 
\begin{equation}\label{eq:synch}
P_{\mbox{\tiny synch}}=\frac{4}{3}\sigma_{\mbox{\tiny T}}c\gamma^2\beta^2 U_{\mbox{\tiny B}} \quad\quad [\mbox{erg s$^{-1}$}]
\end{equation} where $\sigma_{\mbox{\tiny T}}$ is Thompson scattering cross section, $\beta=v/c$, $\gamma=1/\sqrt{1-(v/c)^2}$, $U_{\mbox{\tiny B}}=B^2/8\pi$ is the magnetic field energy density (erg cm$^{-3}$). Power spectrum of synchrotron emission from a CR electron peaks at $x=\nu/\nu_c=0.29$ \citep{ginzburg_and_syrovatskii_1965,rybicki_and_lightman_1979} where $\nu_c$ is the characteristic frequency for CR electron with $E_{\mbox{\tiny kin}}$
\begin{equation}\label{eq:freq_c}
\left(\frac{\nu_c}{\mbox{GHz}}\right) = 1.3\times10^{-2}\left(\frac{B}{\mu\mbox{G}}\right)\left(\frac{E_{\mbox{\tiny kin}}}{\mbox{GeV}}\right)^2
\end{equation}

\subsubsection{Inverse Compton process}
The energy loss rate of a single CR electron propagating through a thermal bath of photons, via the inverse Compton process is 
\begin{equation}\label{eq:ic}
P_{\mbox{\tiny IC}}=\frac{4}{3}\sigma_{\mbox{\tiny T}}c\gamma^2\beta^2 U_{\mbox{\tiny ph}} \quad\quad [\mbox{erg s$^{-1}$}]
\end{equation} where $U_{\mbox{\tiny ph}}$ is the photon energy density (erg cm$^{-3}$) which includes the contribution from both CMB radiation and IR/UV radiation from star formation ($U_{\mbox{\tiny ph}}=U_{\mbox{\tiny CMB}}+U_{\mbox{\tiny rad}}$).

\subsubsection{Bremsstrahlung}
The energy loss rate of a single CR electron interacting with thermal nuclei, via `relativistic' bremsstrahlung emission is given by Bethe-Heitler formula (Equation (6.74) in \cite{longair_2011}) and is expressed as 
\begin{equation}\label{eq:brem}
P_{\mbox{\tiny brem}}=\frac{Z(Z+1.3)e^6 n_p}{16\pi^3 \epsilon_0^3 m_e c^2 \hbar}(\gamma-1)\left[\mbox{ln}\left(\frac{183}{Z^{1/3}}\right) +\frac{1}{8}\right] \quad [\mbox{erg s$^{-1}$}]
\end{equation} where charge number $Z$ is 1 for proton from ionized hydrogen, $n_p$ is the number density of protons, $e$ is the electron charge, and $\epsilon_0$ is the vacuum permittivity. The `relativistic' bremsstrahlung cooling is proportional to the kinetic energy of CR electron (i.e., $E_{\mbox{\tiny kin}}=(\gamma-1)m_e c^2$). 

\subsubsection{Ionization}
The energy loss of relativistic CR electron per unit path length ($dE/dx$) is given by the Bethe-Bloch formula (Equation (5.26) in \cite{longair_2011}). Then the practical form for the energy loss rate of a single CR electron ($dE/dt=c\beta\times dE/dx$) ionizing electrons in atoms via the collisional ionization process  is expressed as 
\begin{eqnarray}\label{eq:ion}
P_{\mbox{\tiny ion}} & = & \frac{c e^4 n_e}{8\pi \epsilon_0^2 m_e c^2}\frac{\gamma}{\sqrt{\gamma^2-1}}\times\\\nonumber
& & \biggl[\mbox{ln}\left\{\frac{1}{2}\left(\frac{m_e c^2}{\bar{I}}\right)^2\right\} + \mbox{ln}\left\{\frac{(\gamma^2-1)^2}{\gamma(\gamma+1)}\right\} \\\nonumber
& & - \mbox{ln}2\left(\frac{2}{\gamma}-\frac{1}{\gamma^2}\right) + \frac{1}{\gamma^2} + \frac{1}{8}\left(1-\frac{1}{\gamma} \right)^2\biggr] \quad [\mbox{erg s$^{-1}$}]
\end{eqnarray} based on Equation (5.36) in \cite{longair_2011} where $n_e$ is the number density of electrons, $e$ is the electron charge, $\epsilon_0$ is the vacuum permittivity, and $\bar{I}$ is the energy of ionized electron including ionization potential such that for hydrogen atom it is $\approx 50$ eV \citep{longair_2011}. The energy loss rate of CR electron via ionization process is inversely proportional to the kinetic energy of CR electron for $E_{\mbox{\tiny kin}}\lesssim m_e c^2$ and slowly increasing proportional to $\mbox{ln} \gamma^2$ for $E_{\mbox{\tiny kin}}\gtrsim m_e c^2$ \citep{longair_2011}.

\subsubsection{Adiabatic expansion}
We consider a uniformly expanding sphere of CR electrons. Then the velocity distribution inside the sphere is $v=v_0(r/R)$ where $v_0$ is the expansion velocity of the outer radius $R$ \citep{longair_2011}. The energy loss rate of a single CR electron confined within an expanding volume during adiabatic expansion is
\begin{eqnarray}\label{eq:ad}
P_{\mbox{\tiny ad}} & = &  \frac{v_0}{R}E_{\mbox{\tiny kin}} \\\nonumber
& = & 2.65\times10^{-20} (\gamma-1) \left(\frac{R}{\mbox{pc}}\right)^{-1}\left(\frac{v_0}{\mbox{km s}^{-1}}\right) \quad [\mbox{erg s$^{-1}$}]
\end{eqnarray} Since $R$ and $v_0$ for CR energy loss is not well constrained, we adopt $R=1$ kpc and $v_0=300$ km s$^{-1}$ from \cite{murphy_2009} and use them in this study. 

\subsubsection{Fraction of synchrotron cooling}
The total power loss rate of a single CR electron by three major radiative processes (synchrotron, IC, and, bremsstrahlung) as well as ionization and adiabatic expansion is
\begin{equation}
P_{\mbox{\tiny CR}}  =  P_{\mbox{\tiny synch}} + P_{\mbox{\tiny IC}} + P_{\mbox{\tiny brem}} + P_{\mbox{\tiny ion}} + P_{\mbox{\tiny ad}} \quad\quad[\mbox{erg s$^{-1}$}]
\end{equation}
Unlike the previous studies \citep{algera_etal_2020,lacki_etal_2010b,murphy_2009,schober_etal_2016} using a combined cooling time scale for each mechanism for CR electron energy loss, we estimate the fraction of synchrotron cooling of CR electrons using the actual energy loss rate (instead of cooling time). 
\begin{equation}
f_{\mbox{\tiny synch}} = \frac{P_{\mbox{\tiny synch}}}{P_{\mbox{\tiny CR}}}
\end{equation}

\subsubsection{Redshift variation of synchrotron fraction}
If a galaxy in the local universe is placed in the high-$z$ universe without changing the intrinsic properties of star formation depending primarily on gas surface density, the fraction of synchrotron cooling relative to other CR electron cooling processes (measured by $f_{\mbox{\tiny synch}}$) deviates from the value established in the local universe by a factor of $f_s$ which is expressed as 
\begin{eqnarray}\label{eq:fscale}
    f_s & = & \frac{f_{\mbox{\tiny synch}}(z)}{f_{\mbox{\tiny synch}}(0)}\\\nonumber
    & = & \frac{P_{\mbox{\tiny synch}}(0) + P_{\mbox{\tiny IC}}(0) + P_{\mbox{\tiny brem}}(0) + P_{\mbox{\tiny ion}}(0) + P_{\mbox{\tiny ad}}(0)}{P_{\mbox{\tiny synch}}(z) + P_{\mbox{\tiny IC}}(z) + P_{\mbox{\tiny brem}}(z) + P_{\mbox{\tiny ion}}(z) + P_{\mbox{\tiny ad}}(z)}\\\nonumber
    & & \times\frac{P_{\mbox{\tiny synch}}(z)}{P_{\mbox{\tiny synch}}(0)}
\end{eqnarray}
Similarly, \cite{algera_etal_2020} presents a toy model of $q_{\mbox{\tiny IR}}$ value for high-$z$ sub-\textit{mm} galaxies based on the use of $q_{\mbox{\tiny IR}}$ value for local ULIRGs as a normalization.

If we consider the increase of the CMB radiation energy density (for IC cooling) with redshift and no redshift dependency for other cooling processes, the $q_{\mbox{\tiny IR}}$ value for the high-$z$ galaxy will increase because of `synchrotron dimming' as suggested in the previous studies \citep[e.g.,][]{lacki_etal_2010b,murphy_2009}, by the factor of $f_s$. 

However, as we discussed earlier, this monotonic increase of $q_{\mbox{\tiny IR}}$ value due to the increased CMB IC cooling of CR electron is not consistent with the observation. In the following sections (Section~\ref{sec:paramgas} and ~\ref{sec:bamp}), we consider the CR electron cooling depending on the gas surface density and the amplification of the turbulent magnetic field in the ISM of high-$z$ galaxies that enhances synchrotron emission.

\subsection{Parameterization by gas surface density}\label{sec:paramgas}
\cite{lacki_etal_2010a} developed a one-zone model of the CR electron transport in galaxies and predicted the radio-IR correlation that depends on the gas surface density. The model incorporates complex physical processes in the computation of radio flux and predicts $q_{\mbox{\tiny IR}}$ for a wide range of model parameter space. The model was applied for high-$z$ galaxies and predicted the $q_{\mbox{\tiny IR}}$ for different gas surface densities and different redshifts \citep{lacki_etal_2010b}, which shows that the $q_{\mbox{\tiny IR}}$ monotonically decreases with gas surface density for a fixed redshift and the $q_{\mbox{\tiny IR}}$ monotonically increases with redshift for a fixed gas surface density (or nearly flat if gas surface density is high, $\Sigma_g > 1$ g cm$^{-2}$). We also note a similar study by \cite{schober_etal_2016} that attempts to compute the cosmic evolution of the radio synchrotron emission from galaxies and the radio-IR correlation.

However, the \cite{lacki_etal_2010b} model and the \cite{schober_etal_2016} model have not been compared with the observed distribution of $q_{\mbox{\tiny IR}}$ value\footnote{\cite{schober_etal_2016} shows a comparison only for $z=0$ galaxy samples.} and the prediction from \cite{lacki_etal_2010b} of a monotonic increase (for normal star forming galaxy with $\Sigma_g\approx 0.1$ g cm$^{-2}$) or a flat evolution (for extremely high gas density star forming galaxy with $\Sigma_g\approx 10$ g cm$^{-2}$) of $q_{\mbox{\tiny IR}}$ with redshift for given $\Sigma_g$ is not consistent with the observed distribution of $q_{\mbox{\tiny IR}}$ (i.e., flat or mildly decreasing $q_{\mbox{\tiny IR}}$ with redshift as discussed in Section~\ref{sec:intro}) as also seen in \cite{schober_etal_2016}, which implies the need for additional enhancement of synchrotron emission in the ISM of high-$z$ galaxies compared to the local galaxies, due to the increased magnetic field density as discussed in Section~\ref{sec:bamp}. 

In this study, we propose a simple parameterized model of $q_{\mbox{\tiny IR}}$ using gas surface density and redshift. In our model, we express the magnetic field energy density and the photon energy density from star formation, using an `average' gas surface density\footnote{1 g cm$^{-2}$ = 4800 $M_{\odot}$ pc$^{-2}$ and 12 $M_{\odot}$ pc$^{-2}$ is for our Milky Way} ($\Sigma_g$ [$M_{\odot}$ pc$^{-2}$]) by following the scaling relations in the previous literature. Here we note that Kennicutt-Schmidt relation used in this work to convert star formation rate surface density into gas surface density is not a universal relationship holding for every galaxies and has several systematic variations in both slope and normalization including scatter \citep[e.g.,][]{leroy_etal_2008}. Nevertheless, for ensemble of galaxies covering a large dynamic range of SFRs, the relation has a relatively small dispersion of $\approx 0.3$ dex \citep[][]{Kennicutt_1998,Wyder_etal_2009} and suggests that the gas surface density is the primary factor in regulating star formation rate surface density \citep{Shi_etal_2018}.   

From Equation (11) in \cite{lacki_etal_2010a}, for optically thick IR emission for high gas surface density ($\Sigma_g \gtrsim 0.1$-$1$ g cm$^{-2}$), the radiation energy density of photons from star formation becomes
\begin{equation}
    U_{\mbox{\tiny rad}} = 2.11\times10^{-14} (\tau_{\mbox{\tiny FIR}}+1) \left(\frac{\Sigma_g}{M_{\odot}\mbox{pc}^{-2}}\right)^{1.4}~[\mbox{erg cm}^{-3}]
\end{equation} where $\tau_{\mbox{\tiny FIR}}=\kappa_{\mbox{\tiny FIR}}\Sigma_{g}/2$ is the vertical optical depth with the Rosseland mean dust opacity $\kappa_{\mbox{\tiny FIR}}$ \citep[1-10 cm$^2$g$^{-1}$ in ][]{lacki_etal_2010a} while, for optically thin IR emission for low gas surface density, the energy density in starlight becomes \citep[from Equation (9) in ][]{lacki_etal_2010a}
\begin{equation}
    U_{\mbox{\tiny rad}} = 2.11\times10^{-14}\left(\frac{\Sigma_g}{M_{\odot}\mbox{pc}^{-2}}\right)^{1.4}~[\mbox{erg cm}^{-3}]
\end{equation}
In our model, we use the density threshold $\Sigma_g = 0.1$ g cm$^{-2}$ to choose $U_{\mbox{\tiny rad}}$ for optically thick or thin IR emission, and we adopt $\kappa_{\mbox{\tiny FIR}}=3$ cm$^2$ g$^{-1}$ in the subsequent analysis. 

For magnetic field energy density ($U_{\mbox{\tiny B}}=B^2/8\pi$), we adopt the magnetic field density following a power-law function of $\Sigma_{\mbox{\tiny SFR}}$ \citep[$B=B_0\Sigma_{\mbox{\tiny SFR}}^{\eta}$ in ][]{manna_and_roy_2023} to have
\begin{eqnarray}\label{eq:ub}
U_{\mbox{\tiny B}} & = & 1\times10^{-12} \times \frac{(2.5\times10^{-4})^{2\eta}}{8\pi}\left(\frac{B_0}{{\mu \mbox{G}}}\right)^2 \nonumber\\ 
& & \times \left( \frac{\Sigma_g}{M_{\odot}\mbox{pc}^{-2}} \right)^{2.8\eta} [\mbox{erg cm}^{-3}]
\end{eqnarray} where we convert $\Sigma_{\mbox{\tiny SFR}}$ to $\Sigma_g$ using Kennicutt-Schmidt relation \citep[$\Sigma_{\mbox{\tiny SFR}}=2.5\times10^{-4} \Sigma_g^{1.4}$ in][]{ks_relation_2012}. In this paper, we adopt $\eta=0.31$ \citep{manna_and_roy_2023}. The normalization of the magnetic field, $B_0$, derived from a small number of local galaxies widely varies (typically $10-100\mu$G) in literature
\citep[e.g.,][]{chyzy_2008,heesen_etal_2023,manna_and_roy_2023,schleicher_and_beck_2013,tabatabaei_etal_2013}. We set it as a free parameter and find that $50-100\mu$G is a reasonable range of the value for given $\Sigma_g$ values used in this paper.  

For CMB energy density, $U_{\mbox{\tiny CMB}}(z)=U_{\mbox{\tiny CMB}}(0)\times(1+z)^4$ where $U_{\mbox{\tiny CMB}}(0)=4.2\times10^{-13}$ erg cm$^{-3}$ \citep{draine_2011}. We note that $U_{\mbox{\tiny rad}}$ does not depend on redshift while $U_{\mbox{\tiny CMB}}$ does and $U_{\mbox{\tiny B}}$ also does if considering the amplification of magnetic field discussed in Section~\ref{sec:bamp}. We use the gas-density-dependent $U_{\mbox{\tiny rad}}$ and $U_{\mbox{\tiny B}}$ and the redshift-depdenent $U_{\mbox{\tiny CMB}}$ for $P_{\mbox{\tiny synch}}$ and $P_{\mbox{\tiny IC}}$ (Equation~\ref{eq:synch} and ~\ref{eq:ic}).

For the energy loss rate of CR electron due to bremsstrahlung cooling (Equation~\ref{eq:brem}), we can rewrite this expression to the following form depending on the gas surface density by assuming a typical line-of-sight length scale ($\ell$).
\begin{equation}
P_{\mbox{\tiny brem}} = 2.82\times10^{-23} \left(\frac{\Sigma_g}{M_{\odot}\mbox{pc}^{-2}}\right) \left(\frac{\mbox{kpc}}{\ell}\right) \quad [\mbox{erg s$^{-1}$}]
\end{equation}

Likewise, for the energy loss rate of CR electron due to the ionization process (Equation~\ref{eq:ion}), we can rewrite this expression to the following form depending on the gas surface density by assuming a typical line-of-sight length scale ($\ell$) and $n_e$ as same as $n_p$.
\begin{equation}
P_{\mbox{\tiny ion}} = 4.96\times10^{-22} \left(\frac{\Sigma_g}{M_{\odot}\mbox{pc}^{-2}}\right) \left(\frac{\mbox{kpc}}{\ell}\right) \quad [\mbox{erg s$^{-1}$}]
\end{equation}

We now have the expressions of CR electron cooling processes depending on the gas surface density that are used in Equation~\ref{eq:fscale}. In our model, we trace the $q_{\mbox{\tiny IR}}$ value for a galaxy with fixed gas surface density along the redshift.

\subsection{Amplification of the turbulent magnetic field}\label{sec:bamp}
If the synchrotron emission becomes faint with increasing redshift (Equation~\ref{eq:fscale}), the radio-IR correlation coefficient $q_{\mbox{\tiny IR}}$ only becomes larger with increasing redshift (see Figure 6 in \cite{murphy_2009}), which is inconsistent with observations: decreased or constant $q_{\mbox{\tiny IR}}$ value for star-forming galaxies with increasing redshift \citep[e.g.,][]{delhaize_etal_2017,delvecchio_etal_2021,ivison_etal_2010,murphy_2009}. However, it is shown that small-scale turbulent dynamo can amplify the seed magnetic field \citep{kazantsev_1968,kulsrud_and_anderson_1992,brandenburg_and_subramanian_2005} and increase the turbulent magnetic field strength through the turbulent energy injection from supernova explosions \citep[e.g.,][]{gent_etal_2023,rieder_etal_2017,pakmor_etal_2024,schober_etal_2013}. If the magnetic field is increased in turbulent ISM, the turbulent magnetic field in high-$z$ galaxies with a strong feedback process is indeed expected to be higher than what one might think. 

Understanding cosmological evolution of the magnetic field in galaxies requires an incorporation of galactic dynamo theory into galaxy formation process (via either semi-analytic model or numerical simulation). Several studies suggests that the typical magnetic field strength \textit{at any given galactic stellar mass} is predicted to increase with redshift \citep{rodrigues_etal_2015,rodrigues_etal_2019}. If the typical gas volume density ($\rho$) in the ISM increases with redshift, one can expect an increase of the magnetic field strength and the radio emission from high-$z$ star-forming galaxies \citep{schleicher_and_beck_2013}. Also, recent observation of dust polarization from a high-$z$ galaxy provides direct evidence of the large-scale (5 kpc) ordered galactic magnetic field \citep{geach_etal_2023}. In this hypothesis of the increased magnetic field in dense turbulent ISM in high-$z$ galaxies with high star formation rate \citep[e.g.,][]{Tabatabaei_etal_2017}, synchrotron dimming due to the increased CMB energy density may become less significant, which is in line with the observation of the `no' variation or `mild' decrease of $q_{\mbox{\tiny IR}}$ value as a function of redshift. 

In Equation~\ref{eq:ub}, we use the magnetic field strength parameterized by star formation rate surface density ($B=B_0\Sigma_{\mbox{\tiny SFR}}^\eta$) in such that the magnetic field increases with increasing $\Sigma_g$, which explains the observation of local galaxies \citep[][]{manna_and_roy_2023}. According to the model proposed by \cite{schleicher_and_beck_2013}, the normalization parameter $B_0$ depends on several ISM parameters such as gas volume density $\rho$, the fraction of turbulent kinetic energy converted into magnetic energy, the injection rate of turbulent supernova energy \citep{manna_and_roy_2023}. Since the gas volume density $\rho$ scales as $(1+z)^3$, if we assume that the turbulence induced by supernova explosion and stellar feedback is the energy source of the dynamo and a fraction of the turbulent kinetic energy is converted into the magnetic field energy, the magnetic field strength scales as $B\sim(1+z)^{1/2}\Sigma_{\mbox{\tiny SFR}}^{1/3}$ \citep{schleicher_and_beck_2013,schober_etal_2016} which is similar to the empirical relation ($B\sim \Sigma_{\mbox{\tiny SFR}}^{0.31}$ at a given redshift, $z$) in \cite{manna_and_roy_2023}. 

In our model, we incorporate the enhancement of turbulent magnetic field strength in high-$z$ galaxies by adopting a scaling of magnetic field with redshift \footnote{We note that, as also pointed out in \cite{schober_etal_2016}, the \cite{lacki_etal_2010a} model does not consider a scaling of gas volume density $\rho$ with redshift. Therefore the magnetic field does not scale with redshift, which is probably the reason for not seeing a decreasing $q_{\mbox{\tiny IR}}$ with redshift in \cite{lacki_etal_2010b}.}, $B\sim(1+z)^{1/2}$. Then the magnetic field energy density at $z$ becomes 
\begin{equation}\label{eq:bfield}
U_{\mbox{\tiny B}}(z) = U_{\mbox{\tiny B}} (1+z)
\end{equation} where $U_{\mbox{\tiny B}}$ is written in Equation~\ref{eq:ub}. We will use $U_{\mbox{\tiny B}}(z)$ in Equation~\ref{eq:synch} to compute $P_{\mbox{\tiny synch}}(z)$ in Equation~\ref{eq:fscale}.

\begin{figure*}
\centering
\includegraphics[width=0.49\textwidth]{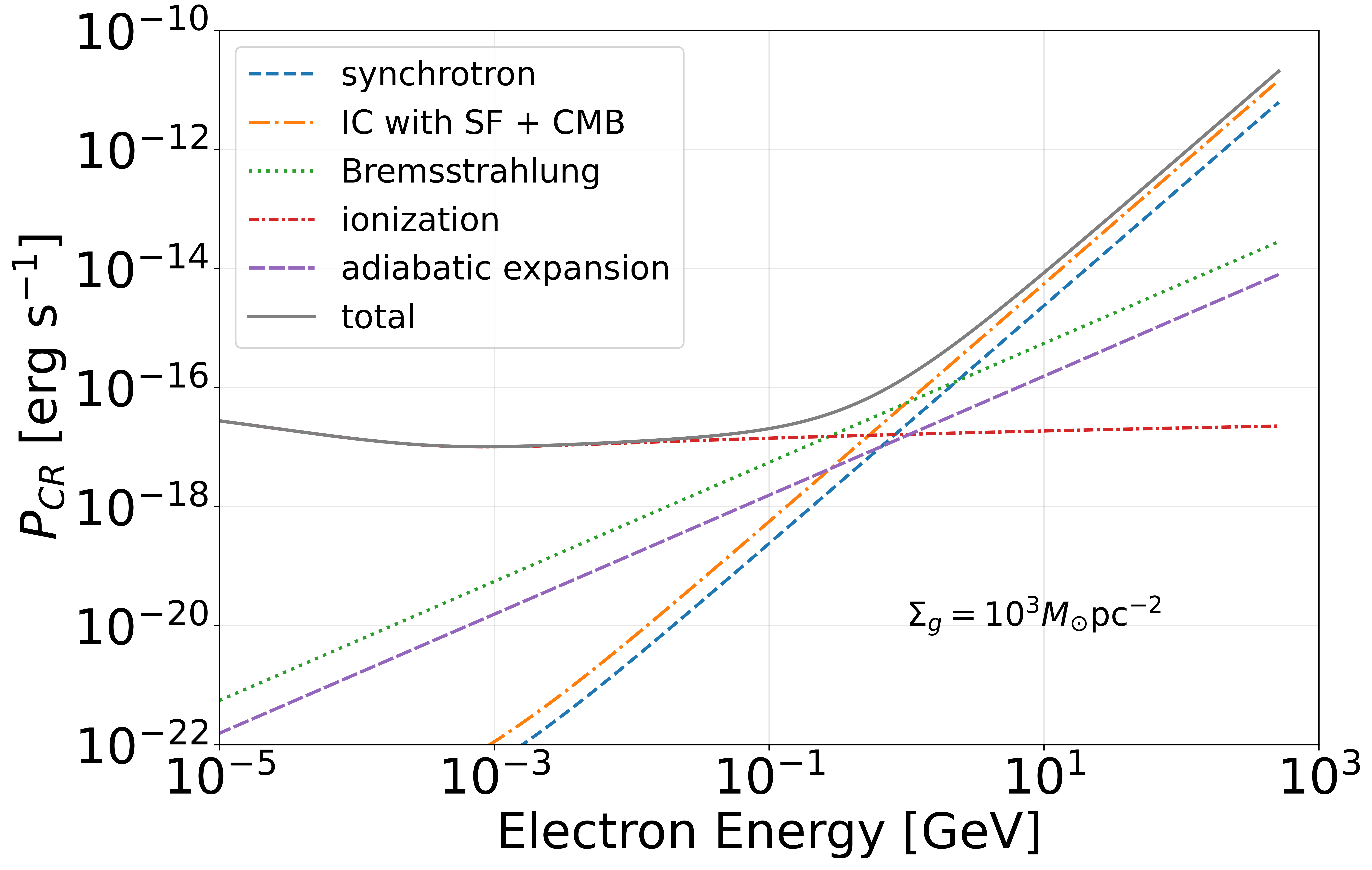}
\includegraphics[width=0.49\textwidth]{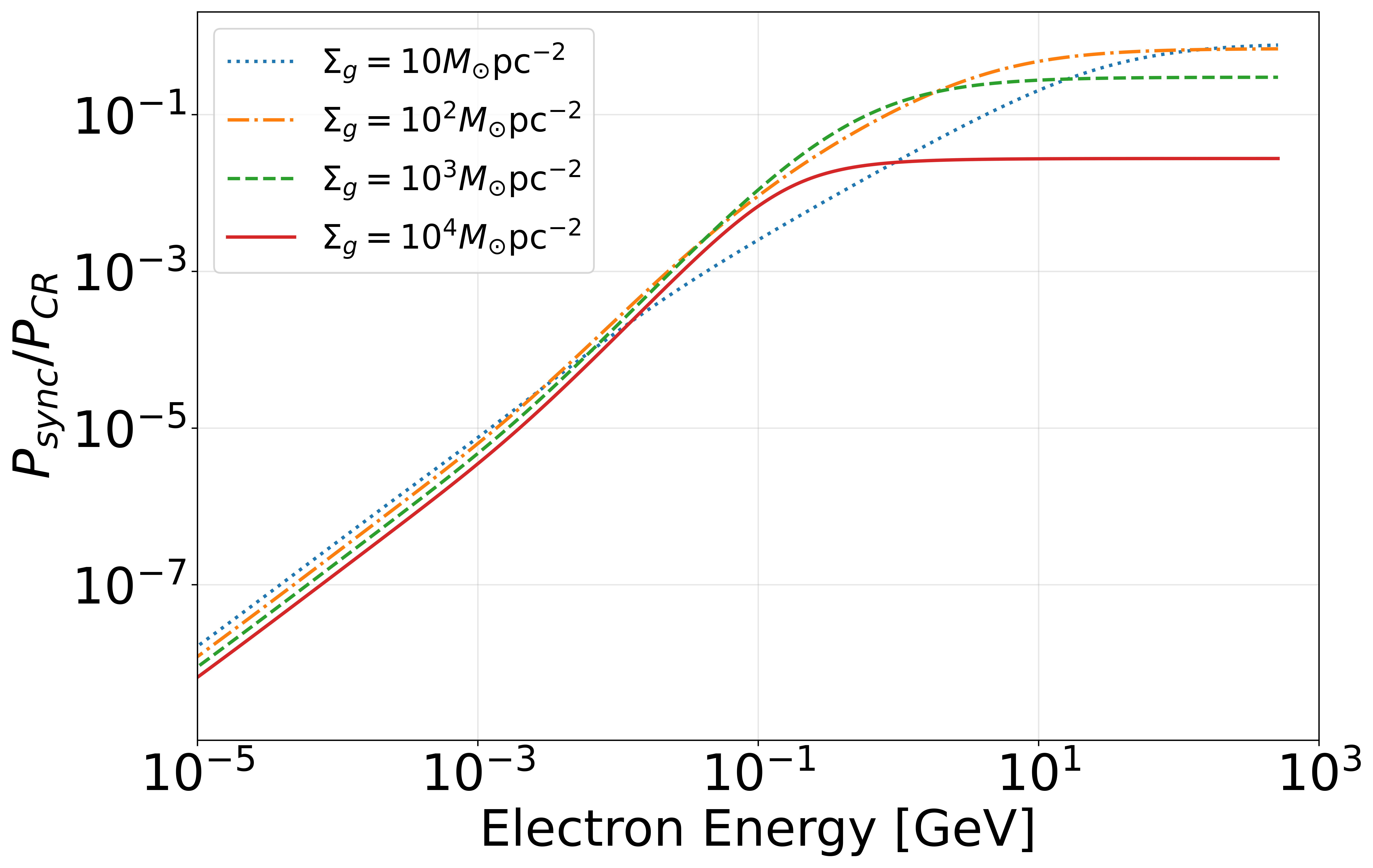}
\caption{\textit{Left: }Energy loss rate of CR electron at $z=0$ as a function of the kinetic energy of CR electron ($E_{\mbox{\tiny kin}}=(\gamma-1)m_e c^2$) for $\Sigma_g=1000~M_{\odot}\mbox{pc}^{-2}$ and $B_0=50 \mu$G \textit{Right: }Fraction of the synchrotron energy loss rate of CR electron at $z=0$ as a function of the kinetic energy of CR electron ($E_{\mbox{\tiny kin}}=(\gamma-1)m_e c^2$) for different $\Sigma_g$ values and $B_0=50 \mu$G}\label{fig:cr_cooling_energy}
\end{figure*}

\begin{figure*}
\centering
\includegraphics[width=0.49\textwidth]{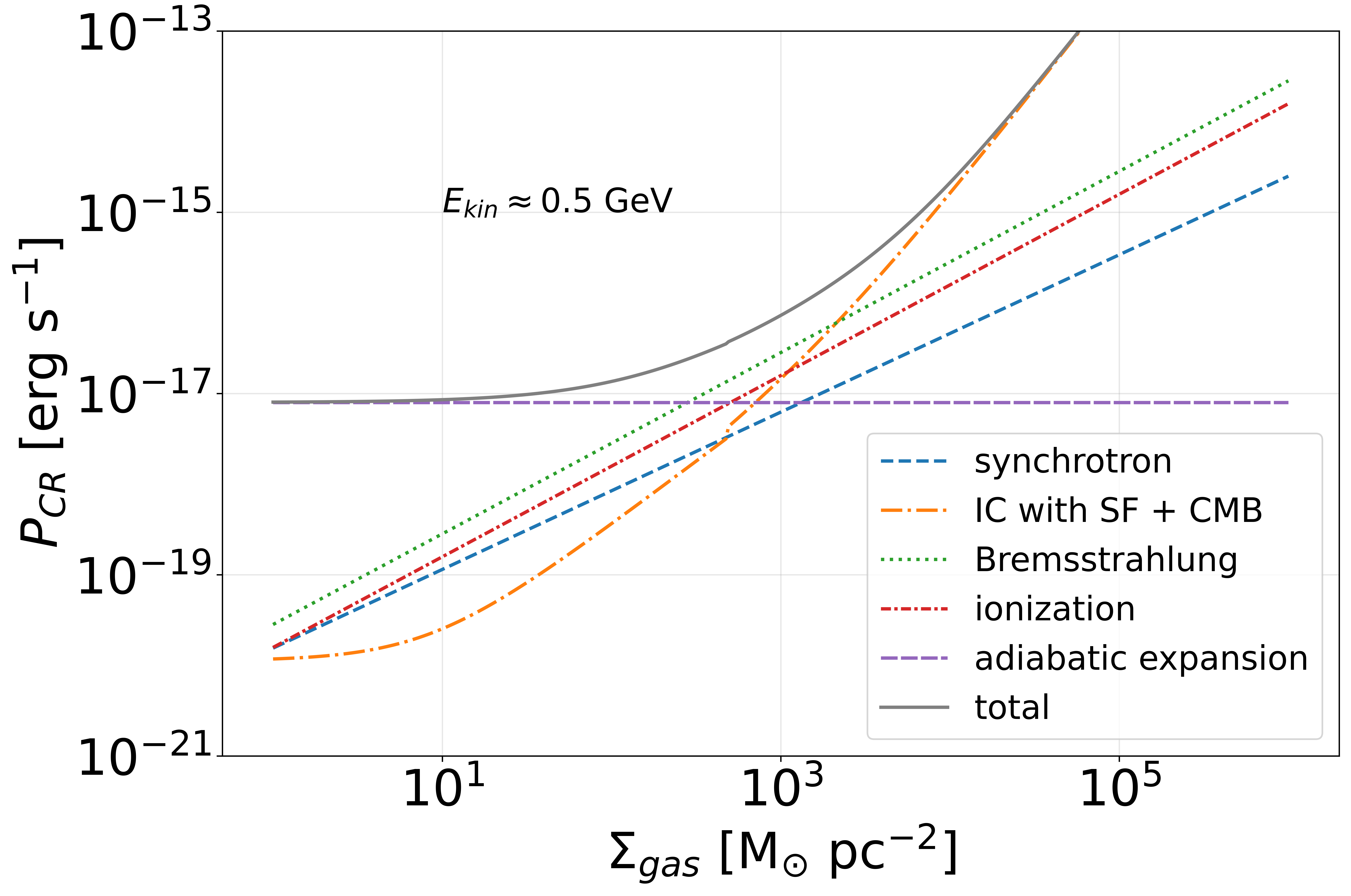}
\includegraphics[width=0.49\textwidth]{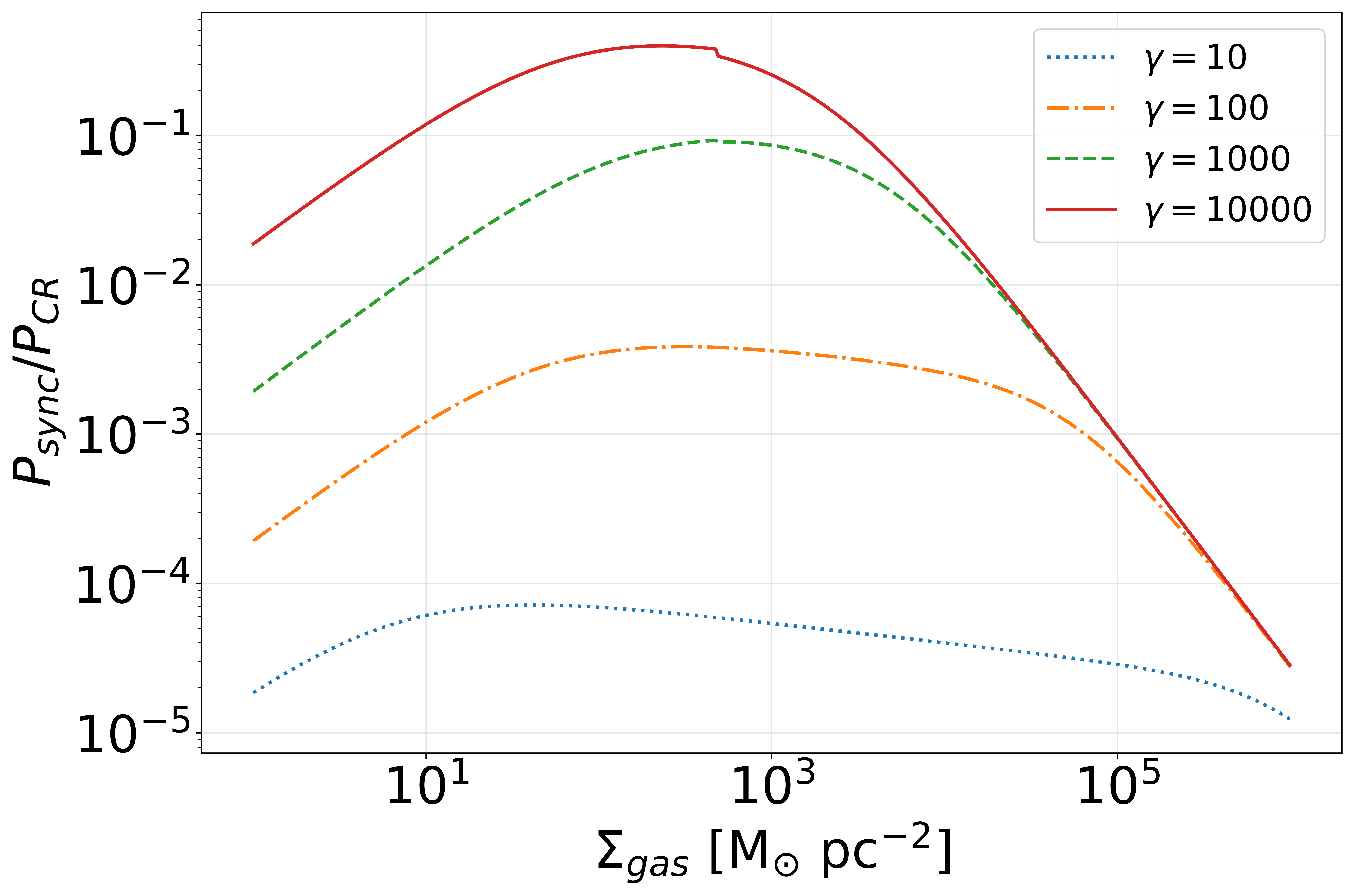}
\caption{\textit{Left: } Energy loss rate of CR electron at $z=0$ as a function of the gas surface density $\Sigma_g$ for $\gamma=10000$ and $B_0=50 \mu$G \textit{Right: }Fraction of the synchrotron energy loss rate of CR electron at $z=0$ as a function of the gas surface density $\Sigma_g$ for different $\gamma$ values and $B_0=50 \mu$G}\label{fig:cr_cooling_den}
\end{figure*}

\subsection{Comparison of the CR electron energy loss rate}
Before we present our model for $q_{\mbox{\tiny IR}}$, the energy loss rates of CR electron at $z=0$ due to the cooling processes we consider and their relative importance are discussed. 

Figure~\ref{fig:cr_cooling_energy} shows the energy loss rate of CR electron for $B_0=50\mu$G due to different cooling mechanisms (left panel) and the fraction of the synchrotron energy loss rate (right panel) as a function of the electron kinetic energy ($E_{\mbox{\tiny kin}}=(\gamma-1)m_e c^2$). In the left panel, we find that, for the gas surface density $\Sigma_g=1000~M_{\odot}\mbox{pc}^{-2}$, high energy CR electron ($E_{\mbox{\tiny kin}}\gtrsim 1$ GeV) cools primarily by IC and synchrotron emission with the characteristic frequency for synchrotron emission $\nu_c \gtrsim 1$ GHz (Equation~\ref{eq:freq_c}) while lower energy CR electron ($E_{\mbox{\tiny kin}}\lesssim 1$ GeV) cools by ionization process. In the right panel, we find that the fraction of synchrotron cooling monotonically decreases with decreasing $E_{\mbox{\tiny kin}}$ for $E_{\mbox{\tiny kin}}\lesssim 1$ GeV and becomes constant for CR electron with $E_{\mbox{\tiny kin}}\gtrsim 1$ GeV. The $E_{\mbox{\tiny kin}}$ value where the transition happens depends on the gas surface density: the transition happens at a lower energy value if the gas surface density is high.

Figure~\ref{fig:cr_cooling_den} shows the energy loss rate of CR electron for $B_0=50\mu$G due to different cooling mechanisms (left panel) and the fraction of the synchrotron energy loss rate (right panel) as a function of gas surface density ($\Sigma_g$). In the left panel, we find that, for CR electron with the electron kinetic energy $E_{\mbox{\tiny kin}}\approx 0.5$ GeV, CR electron in high-density ISM ($\Sigma_g \gtrsim 10^3$M$_{\odot}$pc$^{-2}$) cools primarily by IC while CR electron in low-density ISM ($\Sigma_g \lesssim 10^3$M$_{\odot}$pc$^{-2}$) cools primarily by adiabatic expansion. However, if the adiabatic expansion is not considered, the CR electron cooling in the low-density ISM happens primarily by bremsstrahlung, ionization, and synchrotron emission. In the right panel, we find that the fraction of synchrotron cooling in the highest-density ISM increases with decreasing $\Sigma_g$ and decreases if $\Sigma_g$ is too low. The turn-over happens at $\Sigma_g \lesssim 10^3$M$_{\odot}$pc$^{-2}$ which becomes lower for lower-energy CR electron (lower $\gamma$ value). However, if the adiabatic expansion is not considered, the turn-over does not happen and the fraction of synchrotron cooling increases steeply with decreasing $\Sigma_g$ until $\Sigma_g \approx 10^4$M$_{\odot}$pc$^{-2}$ and mildly increases when $\Sigma_g \lesssim 10^4$M$_{\odot}$pc$^{-2}$.

\subsection{Parametric function of $q_{\mbox{\tiny IR}}$}
Now we can construct a parametric model of $q_{\mbox{\tiny IR}}$ as a function of $\Sigma_g$ and $z$. First, we write the total radio luminosity (including thermal and non-thermal emission) at 1.4GHz from a galaxy at a given redshift $z$
\begin{equation}
L_{1.4GHz}(z) = L^{\mbox{\tiny T}}_{1.4GHz}(z) + L^{\mbox{\tiny NT}}_{1.4GHz}(z).
\end{equation}
Second, given that gas surface density regulates star formation rate surface density of a galaxy, we assume that the thermal radio emission from the galaxy's star formation for a fixed gas surface density does not vary as the galaxy moves along the redshift, and only the non-thermal emission (i.e., synchrotron emission) varies with the redshift as scaled by the synchrotron dimming factor (Equation~\ref{eq:fscale}). Then we can rewrite the above equation as 
\begin{equation}
    L_{1.4GHz}(z) = L^{\mbox{\tiny T}}_{1.4GHz}(0) + L^{\mbox{\tiny NT}}_{1.4GHz}(0) f_s
\end{equation}
Third, we use a thermal fraction $f_{th}=L^{\mbox{\tiny T}}_{1.4GHz}(0)/[ L^{\mbox{\tiny T}}_{1.4GHz}(0) + L^{\mbox{\tiny NT}}_{1.4GHz}(0)]$ at $z=0$ and rewrite the total radio flux density at 1.4GHz.
\begin{equation}
    L_{1.4GHz}(z) = L^{\mbox{\tiny T}}_{1.4GHz}(0) \left[1+f_s\left(\frac{1-f_{th}}{f_{th}} \right)\right]
\end{equation}
Finally, we can rewrite Equation~\ref{eq:qir2} for the power-law spectral index for free-free emission ($F_\nu \sim \nu^{-0.1}$ leading to $L_\nu^T / L^T_{\mbox{\tiny 1.4GHz}}=(\nu/1.4)^{-0.1}$) and the thermal frequency $\nu=33$GHz,
\begin{equation}\label{eq:qirmodel}
    q_{\mbox{\tiny IR}}(z,\Sigma_g)= q_{\mbox{\tiny IR,0}}
    -\mbox{log} \left[ 1+ f_s \left(\frac{1-f_{th}}{f_{th}} \right)\right]
\end{equation} where $q_{\mbox{\tiny IR,0}}=3.65$ for $\nu=33$GHz (Equation~\ref{eq:qir2}) and $f_s$ is expressed as Equation~\ref{eq:fscale}. Here we note that $q_{\mbox{\tiny IR,0}}$ is a free parameter to set a normalization at $z=0$ and $f_s$ carries the dependency of the gas surface density and redshift. 

In Table~\ref{tab:param}, we list the model parameters. In this study, we use $q_{\mbox{\tiny IR,0}}=3.65$ derived from Equation~\ref{eq:qir2} using 33 GHz thermal radio frequency and fix the thermal fraction at 1.4GHz ($f_{th}$) to be 10\% to match the well-constrained value from local galaxies \citep{bell_2003,condon_1992,condon_and_yin_1990,Tabatabaei_etal_2017}. Values of other parameters are assumed if they are not well-determined. In practice, gas surface density ($\Sigma_g$), magnetic field strength ($B_0$), and average Lorentz factor of CR electrons ($\gamma$) are the main parameters that change the $q_{\mbox{\tiny IR}}$ model significantly. In this study, $\Sigma_g$, $B_0$, and $\gamma$ are tuned to capture the observed trend (Figure~\ref{fig:qir_match_z} and \ref{fig:qir_match_den}).

\begin{deluxetable}{lcc}[ht!]
\centering
\tablecaption{Model parameters}\label{tab:param}
\tablehead{\colhead{Parameter} &
           \colhead{description} &
           \colhead{In this work} }
\startdata
$q_{\mbox{\tiny IR,0}}$ & normalization in Equation~\ref{eq:qirmodel}  & 3.65\tablenotemark{$^*$}\\
$f_{th}$ & 1.4GHz thermal fraction at $z=0$ & 0.1\tablenotemark{$^\dag$}\\
$B_0$ & Magnetic field strength & 10-100$\mu$G \\
$\Sigma_g$ & gas surface density & varies \\
$\gamma$ & `average' Lorentz factor of $e_{\mbox{\tiny CR}}$ & $\approx 5000$\tablenotemark{$^{**}$} \\
$\ell$ & line-of-sight length scale of $e_{\mbox{\tiny CR}}$ cloud & 1 kpc\tablenotemark{$^\ddag$}\\
$R$ & radius of $e_{\mbox{\tiny CR}}$ bubble & 1 kpc\tablenotemark{$^\ddag$} \\
$v_0$ & expansion velocity of $e_{\mbox{\tiny CR}}$ bubble & 300 km s$^{-1}$\tablenotemark{$^\ddag$}\\
\enddata
\tablenotetext{$\dag$}{The value is well-constrained in the local universe and therefore fixed in this work.}
\tablenotetext{$\ddag$}{The value is not well determined and assumed in this work. However, the model does not depend very much on the choice of the values.}
\tablenotetext{$*$}{The value is obtained by using $\nu=33$GHz in Equation~\ref{eq:qir2}}
\tablenotetext{$**$}{The corresponding kinetic energy of CR electron is 2.5 GeV.}
\end{deluxetable}

\begin{figure}
\includegraphics[width=0.5\textwidth]{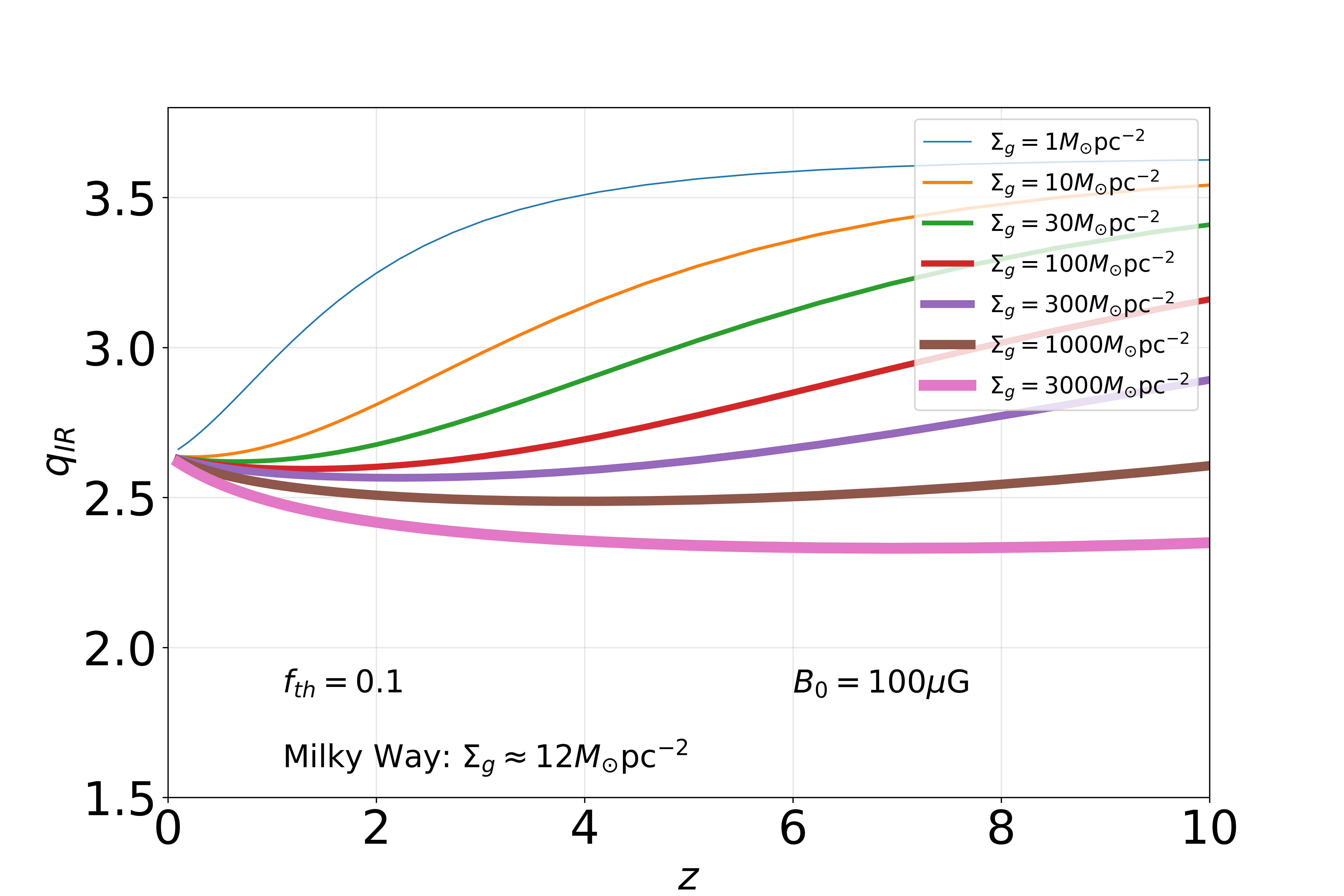}
\caption{$q_{\mbox{\tiny IR}}$ as a function of redshift ($z$) for different gas surface density $\Sigma_g$ values and $B_0=100 \mu$G with an assumed value of the average kinetic energy of CR electron ($E_{\mbox{\tiny kin}}\approx 2.5$GeV or $\gamma=5000$)}\label{fig:qir_model}
\end{figure}

\begin{figure}
\includegraphics[width=0.5\textwidth]{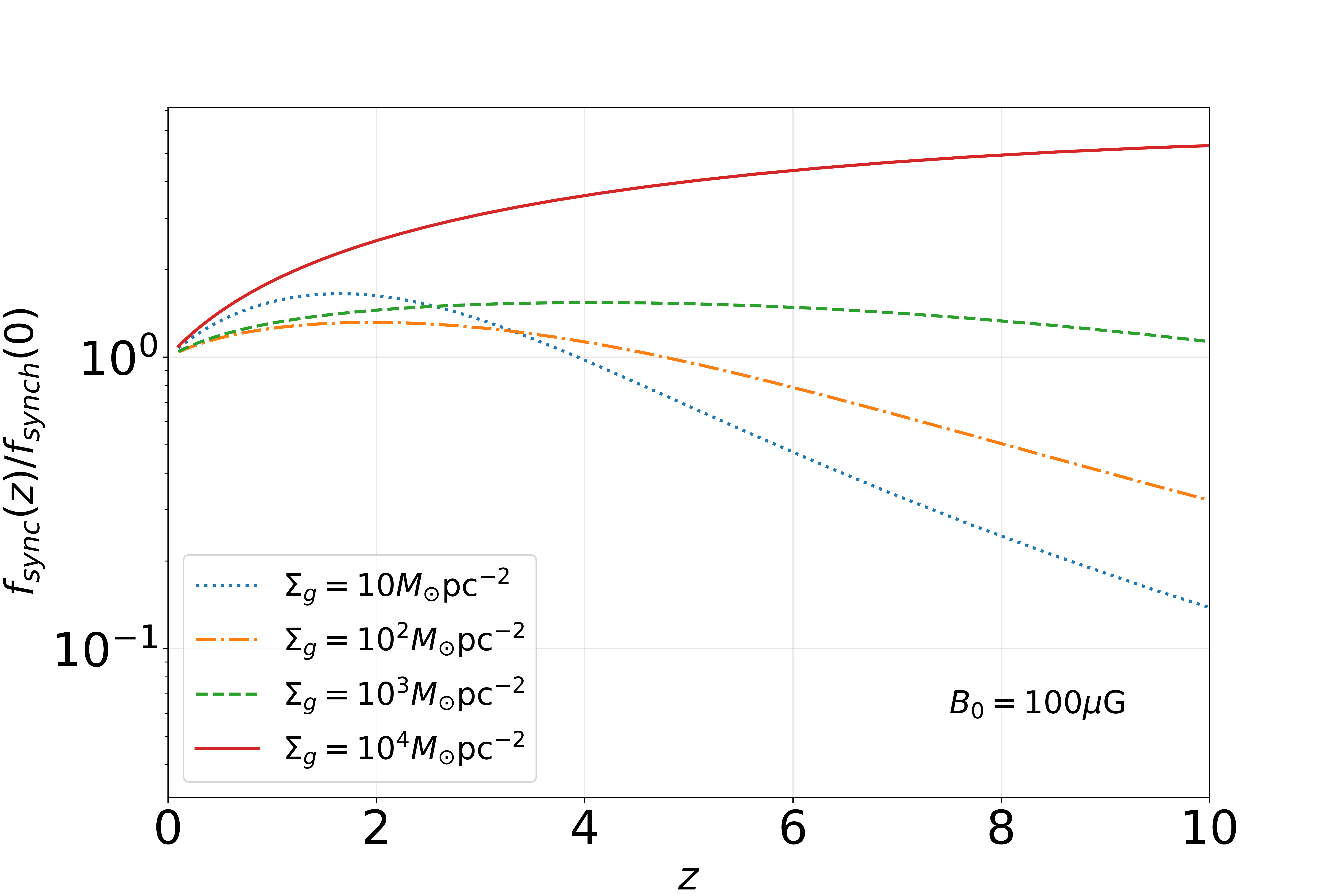}
\caption{Ratio of the fraction of synchrotron energy loss rate between the local ($z=0$) and distant galaxy ISM ($f_s$ factor in Equation~\ref{eq:fscale}), as a function of redshift for different $\Sigma_g$ values with the same $B_0$ and $E_{\mbox{\tiny kin}}$ in Figure~\ref{fig:qir_model}. If the energy loss via synchrotron emission is more efficient (i.e., stronger synchrotron emission) compared to that of local universe, the value is larger than 1.}\label{fig:synch_redshift}
\end{figure}

In Figure~\ref{fig:qir_model}, we show our model of $q_{\mbox{\tiny IR}}$ as a function of redshift ($z$) for different gas surface densities. $B_0$ is set to $100\mu$G and the average kinetic energy of CR electrons is assumed to be 2.5GeV ($\gamma \approx 5000)$. The peak frequency of the power spectrum of synchrotron emission for the corresponding characteristic frequency (Equation~\ref{eq:freq_c}) is $2.35/(1+z)$ GHz which is around 1.4GHz where the radio-infrared correlation is based at. Our model shows a qualitatively different behavior of $q_{\mbox{\tiny IR}}$ value as a function of redshift for different gas surface densities: the low surface density ISM shows a trend of increasing $q_{\mbox{\tiny IR}}$ value with redshift due to the relatively strong effect of synchrotron dimming at high-$z$ while the high surface density ISM shows a very mild variation or even a decrease of $q_{\mbox{\tiny IR}}$ with redshift, which is due to the result of the combination of the energy loss rates from other processes larger than the CMB IC loss and the enhancement of turbulent magnetic field. 

In Figure~\ref{fig:synch_redshift}, we show the ratio of the fraction of synchrotron energy loss rate between local and distant galaxy ISM ($f_s$ factor in Equation~\ref{eq:fscale}) as a function of redshift for four different $\Sigma_g$ values using the same $B_0$ and $E_{\mbox{\tiny kin}}$ value in Figure~\ref{fig:qir_model} to illustrate the effect of the combined effect: density effect and magnetic field enhancement effect. If a galaxy has low $\Sigma_g$, the efficiency of synchrotron cooling becomes less dominant in high redshift universe because of increasingly more efficient CMB IC cooling, and therefore less energy is lost via synchrotron cooling compared to the case in the local universe even though the magnetic field increases with redshift (Equation~\ref{eq:bfield}). However, if the gas surface density of a galaxy is high, the energy loss by synchrotron cooling becomes more efficient in high-$z$ universe, and other channels of energy loss that depend on the gas surface density (IC with the photons from star formation, bremsstrahlung and synchrotron emission) alleviate the impact of CMB IC cooling that is proportional to $(1+z)^4$.  

\section{Comparison with Observation}
\subsection{Observed $q_{\mbox{\tiny IR}}$ values of galaxies}
To compare our model to observation, we need a statistical sample of galaxies for a wide range of redshift with the radio and IR flux measurements. The gas surface density converted from star formation rate surface density based on the radio observation requires the measurement of the size of radio-emitting region. However, there is no deep radio continuum observation with high angular resolution to measure the radio flux and size for many high-$z$ galaxies that is suitable for testing our model. Moreover, since any radio/IR detected galaxy samples for $q_{\mbox{\tiny IR}}$ study have a selection bias, we need a galaxy sample with minimal radio/IR selection bias.

Therefore, in this study, we use one of the largest compilations of galaxies in the COSMOS field with rich multi-wavelength database \citep{delvecchio_etal_2021}. In contrast to the previous studies where galaxies were individually detected at IR or radio wavelength, leading to complex selection functions and biased samples, \cite{delvecchio_etal_2021} selects $\approx 400,000$ star-forming galaxies based on stellar mass, and bins them by stellar mass and redshift, which results in the sample including galaxies both detected and non-detected in radio and IR (see Figure 3 in \cite{delvecchio_etal_2021}). 

As a result, for the galaxies not detected in radio or IR, the radio and IR photometry and the radio size (if not available due to non-detection) are measured by stacking galaxies in each stellar mass and redshift bin \citep{delvecchio_etal_2021}. Also, in \cite{delvecchio_etal_2021}, an adaptive procedure to derive the $q_{\mbox{\tiny IR}}$ threshold is applied for different stellar mass bins to eliminate the AGN-dominated galaxies in the use of the samples to measure the $q_{\mbox{\tiny IR}}$ variation along the redshift. This adaptive procedure inevitably introduces a statistical `impurity': $\approx 30$\% of AGN-dominated galaxies included in the sample and $\approx 4$\% star-forming galaxies excluded from the sample \citep{delvecchio_etal_2021}. For a more detailed procedure for catalog creation with the measurements, we refer the readers to the work by \cite{delvecchio_etal_2021}. We emphasize that the measurements for most galaxies in this catalog are only statistically meaningful. 

We use the $q_{\mbox{\tiny IR}}$ measurements based on the weighted average of the both radio/IR detected and non-detected samples (by stacking) in stellar mass and redshift bins (i.e., data points in Figure 12 in \cite{delvecchio_etal_2021} provided by the author of the paper). For galaxies in each stellar mass and redshift bin, the average star formation rate surface density is measured \citep{delvecchio_etal_2021} and we convert it into the gas surface density, $\Sigma_g$. In following section, we compare our model, $q_{\mbox{\tiny IR}}(z,\Sigma_g)$ to the observed distribution of $q_{\mbox{\tiny IR}}$ measurements in the space of redshift and gas surface density (converted from star formation rate surface density). 

\subsection{Model comparison}
\begin{figure}
\includegraphics[width=0.5\textwidth]{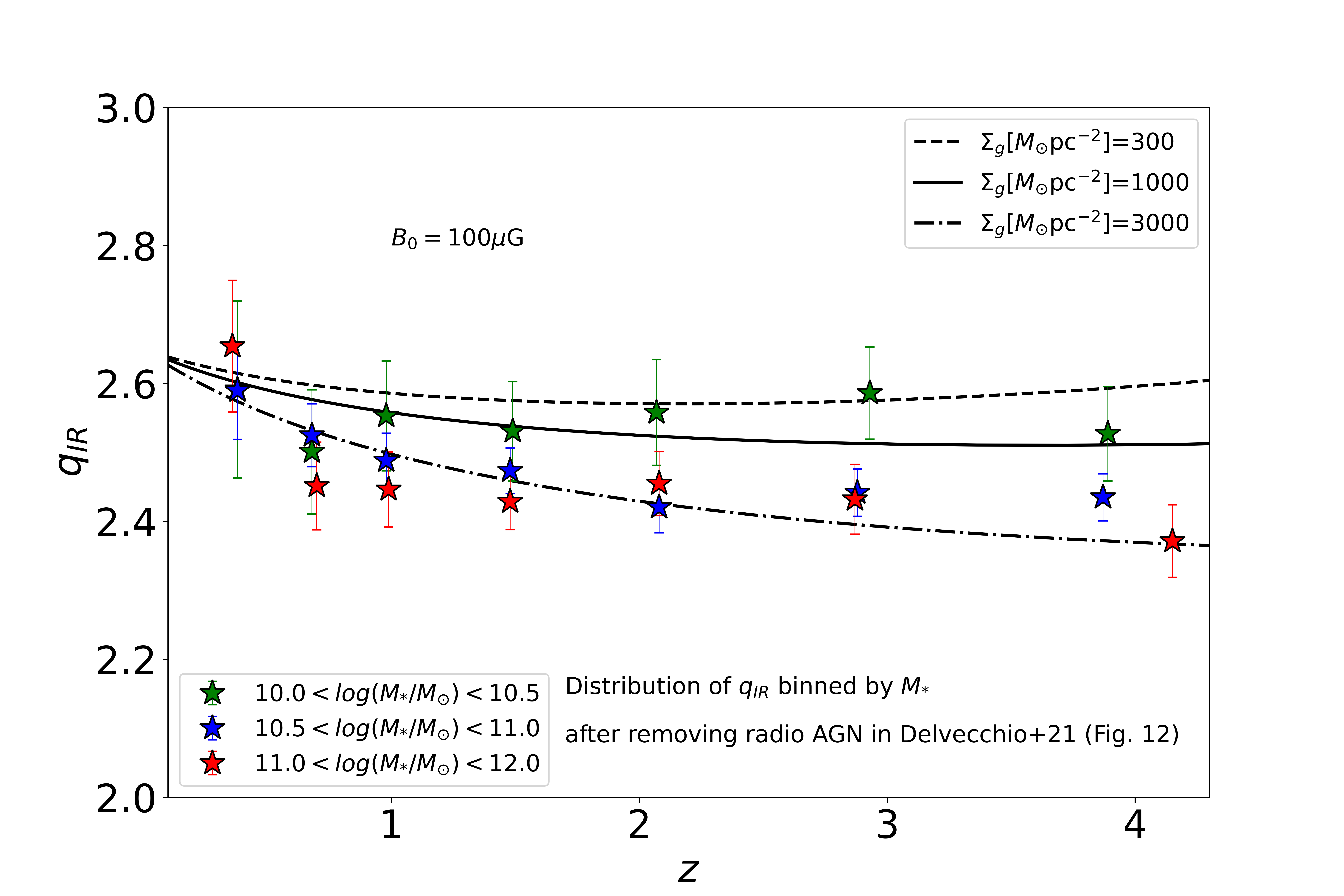}
\caption{Comparison between the measured $q_{\mbox{\tiny IR}}$ values in the redshift space for three most massive stellar mass bins in \cite{delvecchio_etal_2021} and the model based on Equation~\ref{eq:qirmodel}}\label{fig:qir_match_z}
\end{figure}

\begin{figure}
\includegraphics[width=0.5\textwidth]{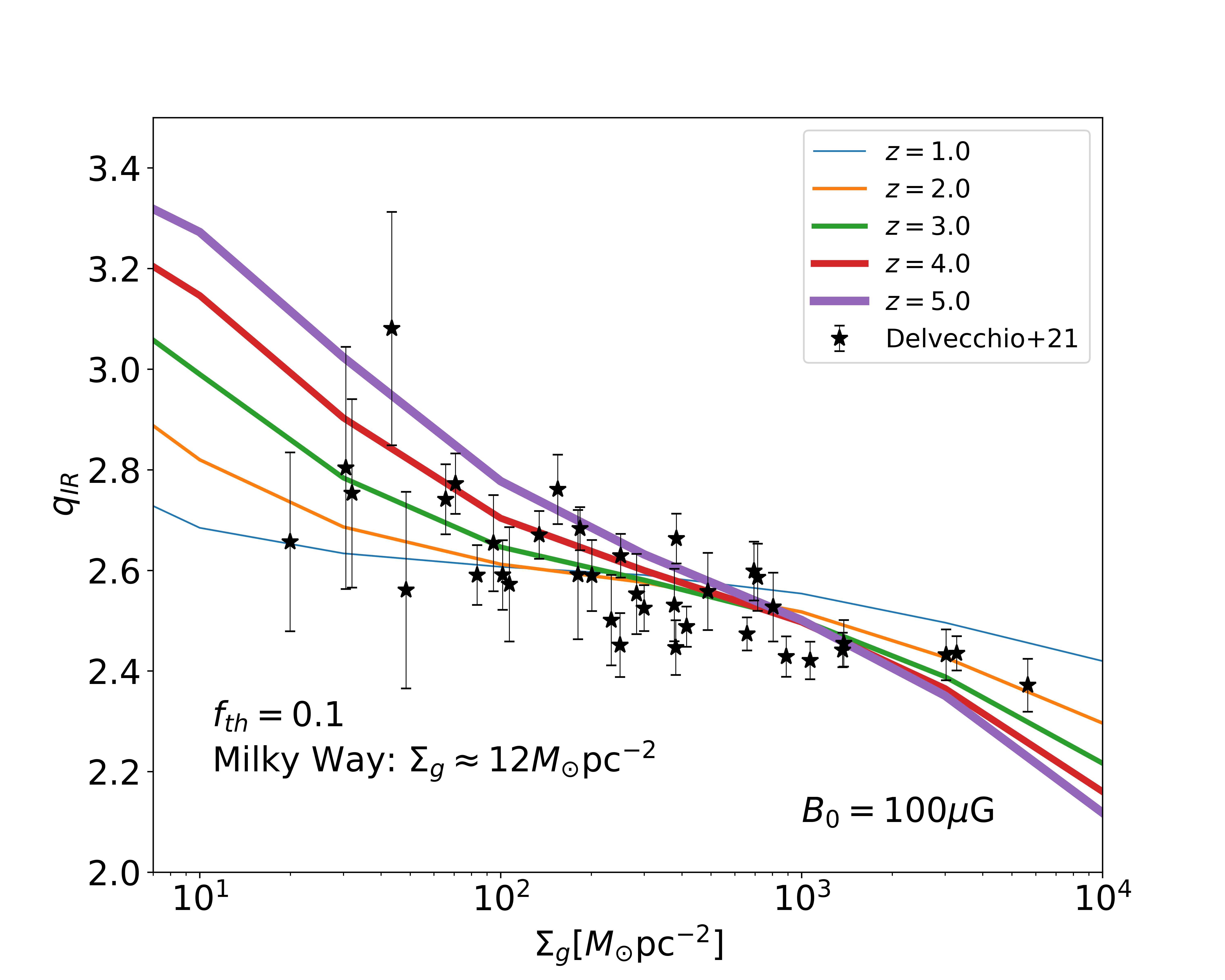}
\caption{Comparison between the measured $q_{\mbox{\tiny IR}}$ values in the gas surface density space in \cite{delvecchio_etal_2021} and the model based on Equation~\ref{eq:qirmodel}}\label{fig:qir_match_den}
\end{figure}

Figure~\ref{fig:qir_match_z} and \ref{fig:qir_match_den} show the comparisons between the observed $q_{\mbox{\tiny IR}}$ values in the redshift and the gas surface density space in \cite{delvecchio_etal_2021} and our $q_{\mbox{\tiny IR}}$ model depending on redshift and gas surface density, $q_{\mbox{\tiny IR}}(z,\Sigma_g)$. 

In Figure~\ref{fig:qir_match_z}, we show the measured $q_{\mbox{\tiny IR}}$ values in the redshift space for three most massive stellar mass bins in Figure 12 of \cite{delvecchio_etal_2021} with enough samples to cover the full redshift range $0.1<z<4.5$, using colored star symbols with error bars. The observed distribution shows a very mild or nearly no variation along the redshift with a noticeable difference of $q_{\mbox{\tiny IR}}$ values between different stellar mass bins \citep{delvecchio_etal_2021}. Three lines with $B_0=100\mu$G and $\gamma=5000$ show the redshift variation of $q_{\mbox{\tiny IR}}(z,\Sigma_g)$ for three different galaxy $\Sigma_g$ values ($300, 1000, 3000 M_{\odot}$pc$^{-2}$). The lines are not a fit to the data and presented to show a trend of the model $q_{\mbox{\tiny IR}}$ that captures a qualitative trend of the observed data. Although \cite{delvecchio_etal_2021} bins the sample by galaxy stellar mass, not by star formation rate surface density, we can infer that more massive galaxies tend to have higher star formation rate surface density because the galaxy size in radio does not change with the galaxy's stellar mass \citep{jimenez-andrade_etal_2021}. For $\gamma$ value of CR electron, we consider $\approx 1$ GeV energy electron which is typically observed from supernova remnant. For wide range of $B_0$ and $\gamma$ values, we find that $B_0=50\sim100 \mu$G and $\gamma=1000 \sim 10000$ allow a reasonable fit ($B_0=100\mu$G and $\gamma=5000$ are used in Figure~\ref{fig:qir_match_z}). 

In Figure~\ref{fig:qir_match_den}, we show the measured $q_{\mbox{\tiny IR}}$ values in the gas surface density space. The observed distribution shows a decrease of $q_{\mbox{\tiny IR}}$ value with increase of gas surface density. Our model $q_{\mbox{\tiny IR}}(z,\Sigma_g)$ with the same parameters used in Figure~\ref{fig:qir_match_z} shown by different colored lines for different redshift values also suggests a systematic decrease of $q_{\mbox{\tiny IR}}$ value with increasing $\Sigma_g$, which captures the observed trend and is also consistent with the prediction from \cite{lacki_etal_2010b}.

Our model captures the distributions of the observed $q_{\mbox{\tiny IR}}$ values for galaxies that are sampled in a wide range of the stellar mass and redshift space and minimally impacted by radio/IR selection effect. We note that the distribution of observed $q_{\mbox{\tiny IR}}$ value is affected by the `impurity' introduced in the exclusion of AGN-dominated galaxies \citep{delvecchio_etal_2021} and the measurements are only statistically meaningful because of stacking radio/IR non-detections. Therefore we cannot perform a rigorous comparison to the observation to make a `quantitative' interpretation of the model parameters. However, the idea of our model for the radio-IR correlation captures the redshift evolution of $q_{\mbox{\tiny IR}}$ value and predicts a systematic decrease of $q_{\mbox{\tiny IR}}$ with increasing $\Sigma_g$. The observed distributions of $q_{\mbox{\tiny IR}}$ measurement from the radio/IR detected galaxies as a function of redshift reported in the current literature suggesting either no systematic variation or a weak systematic trend, is likely to be a representation of the mixture of different sampling of galaxies with different gas surface densities.

\section{Discussion}
\subsection{Observed radio-IR correlation of high-$z$ galaxies}
\cite{lacki_etal_2010a} suggests that the calorimetry (i.e., energy loss entirely due to radiative process) with a certain conspiracy plays a major role in the observed radio-IR correlation. If we assume that the calorimetry theory applies for galaxies in general, an interesting problem is to understand what the main driver is for the observed distribution of the $q_{\mbox{\tiny IR}}$ values from galaxies in low- and high-$z$ universe. 

\cite{delvecchio_etal_2021} shows that the $q_{\mbox{\tiny IR}}$ changes systematically with the galaxy's stellar mass (i.e., the larger the galaxy's stellar mass is, the lower the $q_{\mbox{\tiny IR}}$ is) and with the star formation rate surface density (the larger the star formation rate surface density is, the lower the $q_{\mbox{\tiny IR}}$ is) without a clear signature of the redshift dependence. This result is probably due to the correlation between the galaxy's stellar mass and the star formation rate \citep{delvecchio_etal_2021} with no significant impact from the galaxy size in radio which is nearly independent of the galaxy's stellar mass \citep{jimenez-andrade_etal_2021}. 

The galaxy samples in \cite{delvecchio_etal_2021} have a reliable estimate of the stellar mass based on a rich multiwavelength dataset, and therefore their finding of the stellar mass-dependent $q_{\mbox{\tiny IR}}$ value is likely to be robust. However, given that the radio-IR correlation is directly related to the star formation rate, a more relevant parameter is the star formation rate surface density (or the gas surface density), which is more difficult to measure than the stellar mass because of the uncertainty of galaxy size measurement and the systematic uncertainties in the SFR measurement. 

In general, the observations of radio and IR emission using a finite beam size are biased toward a more compact (or more massive if the galaxy sizes are similar) system which tends to be preferentially detected in the high-$z$ universe. Indeed, \cite{delvecchio_etal_2021} re-analyzed the radio-IR joint selected sample in \cite{delhaize_etal_2017} suggesting the redshift evolution of $q_{\mbox{\tiny IR}}$ up to $z\sim 6$ and showed that the high-$z$ galaxies with the decreased $q_{\mbox{\tiny IR}}$ value have a systematically higher stellar mass with increasing redshift (see Figure D.2 in \cite{delvecchio_etal_2021}). Also, \cite{algera_etal_2020} shows that $q_{\mbox{\tiny IR}}$ value is low for compact and massive sub-millimeter galaxies at $1.5<z<3.5$.

The radio/IR-selected samples for the investigation of the radio-IR correlation in high-$z$ universe have inevitable bias unlike the samples from the radio/IR independent selection (i.e., stellar mass selection) because the radio-IR selected galaxy samples in each redshift bin are systematically different in terms of their star formation rate surface density. Therefore we need to consider the selection effect very carefully for high-$z$ galaxies to interpret the observed radio-IR correlation. In this context, our simple model for the $q_{\mbox{\tiny IR}}$ parameterized by gas surface density and redshift can provide a reasonably good description of the observed $q_{\mbox{\tiny IR}}$ distribution in redshift and gas surface density.

\subsection{Radio spectral index}
In our model, we compare the cooling efficiency of a single CR electron in a cloud of CR electrons injected into the ISM with injection spectrum (i.e., distribution of number density per energy). The shape of CR electron injection spectrum is linked to the non-thermal radio spectral index which, together with thermal free-free spectral index ($\alpha_{th}=-0.1$), determines the total radio spectral index as a superposition of two power-laws: thermal and non-thermal power-law spectrum \citep{klein_etal_2018}. The non-thermal radio spectral index could change from galaxy to galaxy, depending on the balance between the young CR particles injected in star forming region and those cooled and aged in each galaxy \citep{basu_etal_2015a,Tabatabaei_etal_2017}. 

Our model does not adopt radio spectral index as model parameter. However, one can infer the radio spectral index by varying thermal fraction parameter $f_{th}$. From Equation~\ref{eq:qir3} and \ref{eq:qir2} with the assumption of insignificant contribution of synchrotron emission beyond 10GHz \citep{klein_etal_2018}, one can infer the total radio spectral index, $\alpha$ (superposition of thermal and non-thermal power-law slope) over 33-1.4GHz frequency range as
\begin{equation}\label{eq:specindex}
\frac{L_{\mbox{\tiny 33GHz}}^T}{L_{\mbox{\tiny 1.4GHz}}} = \left(\frac{33}{1.4}\right)^{-\alpha} = f_{th}\times\frac{L_{\mbox{\tiny 33GHz}}^T}{L_{\mbox{\tiny 1.4GHz}}^T} = f_{th}\left(\frac{33}{1.4}\right)^{-0.1}.
\end{equation} 
We find that for $f_{th}=0.1$, $\alpha=0.82$ that is consistent with the value ($0.79\pm0.15$) obtained from the 1-10 GHz radio SEDs for 61 nearby galaxies\citep{Tabatabaei_etal_2017} and also similar to the value (0.75) assumed in \cite{delvecchio_etal_2021} to estimate the rest-frame 1.4GHz radio luminosity. As thermal fraction decreases, the radio spectral shape becomes steeper (i.e., larger $\alpha$ in Equation~\ref{eq:specindex}).

Both theory and observation suggest that radio spectral shape becomes steeper at low gas surface density \citep[e.g.,][]{lacki_etal_2010a} and at low star formation rate surface density \citep[e.g.,][]{Tabatabaei_etal_2017}. It is probably because CR electron population is younger and more energetic in galaxies with high star formation activity \citep{Tabatabaei_etal_2017} and therefore less affected by spectral `aging' that produces a steep spectral slope. Also a resolved study in M33 \citep{tabatabaei_etal_2007} and NGC6946 \citep{tabatabaei_etal_2013} respectively shows a flatter non-thermal spectrum in star forming regions than in the diffuse ISM. For diffuse star forming region with low gas density and preferentially older CR electrons, thermal fraction is likely to be lower because non-thermal synchrotron emission has propagated much larger scale than \hii~region and occupies much larger volume, dominating radio emission. This is in line with the inferred radio spectral index from our model: the smaller the thermal fraction is, the steeper the radio spectral shape is.

\subsection{Prospect in the era of the next generation radio telescopes}
Superb sensitivity and angular resolution of the next-generation radio telescopes (ngVLA and SKA) will revolutionize the observation of radio emission from star-forming regions in high-$z$ galaxies by observing the rest frame radio and FIR continuum emission up to $z \sim10$ or larger (Yoon et al. in preparation). The relatively larger field of view of these next-generation radio telescopes compared to the high-frequency telescope (ALMA) looking at the FIR dust emission, enables a deep radio survey of high-$z$ star-forming galaxies which motivates the millimeter follow-up observations using ALMA. 

One of the conventional ways of predicting the expected FIR emission for observing proposals is to use the radio-IR correlation, which we often assume to be universal. A simple and good proxy for the radio-IR correlation as a function of galaxy property is a useful tool not only to estimate the expected flux of IR emission for ALMA observing proposal but also to make a panchromatic galaxy SED from the UV/optical to the radio wavelength based on a self-consistent prescription of the galaxy properties (e.g., stellar population, IMF, dust extinction, and star formation) and the ISM environment (e.g., CMB and magnetic field) around the high-$z$ star-forming regions.  

\section{Summary}
We propose a simple parametric model of the radio-IR correlation, $q_{\mbox{\tiny IR}}$ that depends on the gas surface density and redshift. Use of the magnetic field, radiation field, and CMB energy density parameterized by gas surface density scaled differently with redshift, creates a set of $q_{\mbox{\tiny IR}}$ values as a function of redshift at a given gas surface density of a galaxy. 

Our simple model captures the observed distribution of $q_{\mbox{\tiny IR}}$ values for the `radio/IR' unbiased galaxies sampled in a wide range of stellar mass and redshift bins, with a small number of free parameters: gas surface density ($\Sigma_g$), magnetic field strength ($B_0$) and average kinetic energy of CR electrons ($\gamma$). This study suggests that (1) the very weak evolution of the observed redshift distribution of $q_{\mbox{\tiny IR}}$ values is a result of the compensation of the CMB IC cooling by increased efficiency of synchrotron cooling due to the enhanced turbulent magnetic field and (2) galaxies with higher gas surface density (therefore higher star formation rate surface density for higher chance of being detected) have lower $q_{\mbox{\tiny IR}}$ value, which implies that the radio emission at 1.4GHz compared to the IR emission is relatively brighter for the galaxies with high gas surface density. 

Our model can be a useful tool to simulate the expected radio or IR flux density for high-$z$ galaxies for the given IR or radio measurement and help us understand the observational bias of the radio observation of star-forming galaxies in high-$z$ universe.  

\begin{acknowledgements}
    The author thanks the anonymous referee for careful and constructive feedback that greatly improves the paper. The author acknowledges the kind help from Ivan Delvecchio for providing the data used in this study and useful comments on the analysis of the data. The author also thanks Eric Murphy, Chris Carilli, and Min Yun for their comments on the early draft of the manuscript, and Hansung Gim for a list of useful references for the observational study of radio-IR correlation.   
\end{acknowledgements}

\bibliography{main}{}

\begin{thebibliography}{}
\expandafter\ifx\csname natexlab\endcsname\relax\def\natexlab#1{#1}\fi
\providecommand{\url}[1]{\href{#1}{#1}}
\providecommand{\dodoi}[1]{doi:~\href{http://doi.org/#1}{\nolinkurl{#1}}}
\providecommand{\doeprint}[1]{\href{http://ascl.net/#1}{\nolinkurl{http://ascl.net/#1}}}
\providecommand{\doarXiv}[1]{\href{https://arxiv.org/abs/#1}{\nolinkurl{https://arxiv.org/abs/#1}}}

\bibitem[{{Algera} {et~al.}(2020){Algera}, {Smail}, {Dudzevi{\v{c}}i{\={u}}t{\.{e}}}, {Swinbank}, {Stach}, {Hodge}, {Thomson}, {Almaini}, {Arumugam}, {Blain}, {Calistro-Rivera}, {Chapman}, {Chen}, {da Cunha}, {Farrah}, {Leslie}, {Scott}, {van der Vlugt}, {Wardlow}, \& {van der Werf}}]{algera_etal_2020}
{Algera}, H.~S.~B., {Smail}, I., {Dudzevi{\v{c}}i{\={u}}t{\.{e}}}, U., {et~al.} 2020, \apj, 903, 138, \dodoi{10.3847/1538-4357/abb77b}

\bibitem[{{Basu} {et~al.}(2015{\natexlab{a}}){Basu}, {Beck}, {Schmidt}, \& {Roy}}]{basu_etal_2015a}
{Basu}, A., {Beck}, R., {Schmidt}, P., \& {Roy}, S. 2015{\natexlab{a}}, \mnras, 449, 3879, \dodoi{10.1093/mnras/stv510}

\bibitem[{{Basu} {et~al.}(2015{\natexlab{b}}){Basu}, {Wadadekar}, {Beelen}, {Singh}, {Archana}, {Sirothia}, \& {Ishwara-Chandra}}]{basu_etal_2015b}
{Basu}, A., {Wadadekar}, Y., {Beelen}, A., {et~al.} 2015{\natexlab{b}}, \apj, 803, 51, \dodoi{10.1088/0004-637X/803/2/51}

\bibitem[{{Bell}(2003)}]{bell_2003}
{Bell}, E.~F. 2003, \apj, 586, 794, \dodoi{10.1086/367829}

\bibitem[{{Bourne} {et~al.}(2011){Bourne}, {Dunne}, {Ivison}, {Maddox}, {Dickinson}, \& {Frayer}}]{bourne_etal_2011}
{Bourne}, N., {Dunne}, L., {Ivison}, R.~J., {et~al.} 2011, \mnras, 410, 1155, \dodoi{10.1111/j.1365-2966.2010.17517.x}

\bibitem[{{Brandenburg} \& {Subramanian}(2005)}]{brandenburg_and_subramanian_2005}
{Brandenburg}, A., \& {Subramanian}, K. 2005, \physrep, 417, 1, \dodoi{10.1016/j.physrep.2005.06.005}

\bibitem[{{Calistro Rivera} {et~al.}(2017){Calistro Rivera}, {Williams}, {Hardcastle}, {Duncan}, {R{\"o}ttgering}, {Best}, {Br{\"u}ggen}, {Chy{\.z}y}, {Conselice}, {de Gasperin}, {Engels}, {G{\"u}rkan}, {Intema}, {Jarvis}, {Mahony}, {Miley}, {Morabito}, {Prandoni}, {Sabater}, {Smith}, {Tasse}, {van der Werf}, \& {White}}]{calistro_rivera_etal_2017}
{Calistro Rivera}, G., {Williams}, W.~L., {Hardcastle}, M.~J., {et~al.} 2017, \mnras, 469, 3468, \dodoi{10.1093/mnras/stx1040}

\bibitem[{{Calzetti}(2013)}]{calzetti_2013}
{Calzetti}, D. 2013, in Secular Evolution of Galaxies, ed. J.~{Falc{\'o}n-Barroso} \& J.~H. {Knapen}, 419, \dodoi{10.48550/arXiv.1208.2997}

\bibitem[{{Chy{\.z}y}(2008)}]{chyzy_2008}
{Chy{\.z}y}, K.~T. 2008, \aap, 482, 755, \dodoi{10.1051/0004-6361:20078688}

\bibitem[{{Condon}(1992)}]{condon_1992}
{Condon}, J.~J. 1992, \araa, 30, 575, \dodoi{10.1146/annurev.aa.30.090192.003043}

\bibitem[{{Condon} \& {Yin}(1990)}]{condon_and_yin_1990}
{Condon}, J.~J., \& {Yin}, Q.~F. 1990, \apj, 357, 97, \dodoi{10.1086/168894}

\bibitem[{{Delhaize} {et~al.}(2017){Delhaize}, {Smol{\v{c}}i{\'c}}, {Delvecchio}, {Novak}, {Sargent}, {Baran}, {Magnelli}, {Zamorani}, {Schinnerer}, {Murphy}, {Aravena}, {Berta}, {Bondi}, {Capak}, {Carilli}, {Ciliegi}, {Civano}, {Ilbert}, {Karim}, {Laigle}, {Le F{\`e}vre}, {Marchesi}, {McCracken}, {Salvato}, {Seymour}, \& {Tasca}}]{delhaize_etal_2017}
{Delhaize}, J., {Smol{\v{c}}i{\'c}}, V., {Delvecchio}, I., {et~al.} 2017, \aap, 602, A4, \dodoi{10.1051/0004-6361/201629430}

\bibitem[{{Delvecchio} {et~al.}(2021){Delvecchio}, {Daddi}, {Sargent}, {Jarvis}, {Elbaz}, {Jin}, {Liu}, {Whittam}, {Algera}, {Carraro}, {D'Eugenio}, {Delhaize}, {Kalita}, {Leslie}, {Moln{\'a}r}, {Novak}, {Prandoni}, {Smol{\v{c}}i{\'c}}, {Ao}, {Aravena}, {Bournaud}, {Collier}, {Randriamampandry}, {Randriamanakoto}, {Rodighiero}, {Schober}, {White}, \& {Zamorani}}]{delvecchio_etal_2021}
{Delvecchio}, I., {Daddi}, E., {Sargent}, M.~T., {et~al.} 2021, \aap, 647, A123, \dodoi{10.1051/0004-6361/202039647}

\bibitem[{{Draine}(2011)}]{draine_2011}
{Draine}, B.~T. 2011, {Physics of the Interstellar and Intergalactic Medium}

\bibitem[{{Geach} {et~al.}(2023){Geach}, {Lopez-Rodriguez}, {Doherty}, {Chen}, {Ivison}, {Bendo}, {Dye}, \& {Coppin}}]{geach_etal_2023}
{Geach}, J.~E., {Lopez-Rodriguez}, E., {Doherty}, M.~J., {et~al.} 2023, Nature, arXiv:2309.02034, \dodoi{10.48550/arXiv.2309.02034}

\bibitem[{{Gent} {et~al.}(2023){Gent}, {Mac Low}, {Korpi-Lagg}, \& {Singh}}]{gent_etal_2023}
{Gent}, F.~A., {Mac Low}, M.-M., {Korpi-Lagg}, M.~J., \& {Singh}, N.~K. 2023, \apj, 943, 176, \dodoi{10.3847/1538-4357/acac20}

\bibitem[{{Ginzburg} \& {Syrovatskii}(1965)}]{ginzburg_and_syrovatskii_1965}
{Ginzburg}, V.~L., \& {Syrovatskii}, S.~I. 1965, \araa, 3, 297, \dodoi{10.1146/annurev.aa.03.090165.001501}

\bibitem[{{Heesen} {et~al.}(2023){Heesen}, {Klocke}, {Br{\"u}ggen}, {Tabatabaei}, {Basu}, {Beck}, {Drabent}, {Nikiel-Wroczy{\'n}ski}, {Paladino}, {Schulz}, \& {Stein}}]{heesen_etal_2023}
{Heesen}, V., {Klocke}, T.~L., {Br{\"u}ggen}, M., {et~al.} 2023, \aap, 669, A8, \dodoi{10.1051/0004-6361/202243328}

\bibitem[{{Helou} {et~al.}(1985){Helou}, {Soifer}, \& {Rowan-Robinson}}]{helou_etal_1985}
{Helou}, G., {Soifer}, B.~T., \& {Rowan-Robinson}, M. 1985, \apjl, 298, L7, \dodoi{10.1086/184556}

\bibitem[{{Ivison} {et~al.}(2010){Ivison}, {Alexander}, {Biggs}, {Brandt}, {Chapin}, {Coppin}, {Devlin}, {Dickinson}, {Dunlop}, {Dye}, {Eales}, {Frayer}, {Halpern}, {Hughes}, {Ibar}, {Kov{\'a}cs}, {Marsden}, {Moncelsi}, {Netterfield}, {Pascale}, {Patanchon}, {Rafferty}, {Rex}, {Schinnerer}, {Scott}, {Semisch}, {Smail}, {Swinbank}, {Truch}, {Tucker}, {Viero}, {Walter}, {Wei{\ss}}, {Wiebe}, \& {Xue}}]{ivison_etal_2010}
{Ivison}, R.~J., {Alexander}, D.~M., {Biggs}, A.~D., {et~al.} 2010, \mnras, 402, 245, \dodoi{10.1111/j.1365-2966.2009.15918.x}

\bibitem[{{Jarvis} {et~al.}(2010){Jarvis}, {Smith}, {Bonfield}, {Hardcastle}, {Falder}, {Stevens}, {Ivison}, {Auld}, {Baes}, {Baldry}, {Bamford}, {Bourne}, {Buttiglione}, {Cava}, {Cooray}, {Dariush}, {de Zotti}, {Dunlop}, {Dunne}, {Dye}, {Eales}, {Fritz}, {Hill}, {Hopwood}, {Hughes}, {Ibar}, {Jones}, {Kelvin}, {Lawrence}, {Leeuw}, {Loveday}, {Maddox}, {Micha{\l}owski}, {Negrello}, {Norberg}, {Pohlen}, {Prescott}, {Rigby}, {Robotham}, {Rodighiero}, {Scott}, {Sharp}, {Temi}, {Thompson}, {van der Werf}, {van Kampen}, {Vlahakis}, \& {White}}]{jarvis_etal_2010}
{Jarvis}, M.~J., {Smith}, D.~J.~B., {Bonfield}, D.~G., {et~al.} 2010, \mnras, 409, 92, \dodoi{10.1111/j.1365-2966.2010.17772.x}

\bibitem[{{Jim{\'e}nez-Andrade} {et~al.}(2021){Jim{\'e}nez-Andrade}, {Murphy}, {Heywood}, {Smail}, {Penner}, {Momjian}, {Dickinson}, {Armus}, \& {Lazio}}]{jimenez-andrade_etal_2021}
{Jim{\'e}nez-Andrade}, E.~F., {Murphy}, E.~J., {Heywood}, I., {et~al.} 2021, \apj, 910, 106, \dodoi{10.3847/1538-4357/abe876}

\bibitem[{{Kazantsev}(1968)}]{kazantsev_1968}
{Kazantsev}, A.~P. 1968, Soviet Journal of Experimental and Theoretical Physics, 26, 1031

\bibitem[{{Kennicutt}(1998)}]{Kennicutt_1998}
{Kennicutt}, Robert~C., J. 1998, \araa, 36, 189, \dodoi{10.1146/annurev.astro.36.1.189}

\bibitem[{{Kennicutt} \& {Evans}(2012)}]{ks_relation_2012}
{Kennicutt}, R.~C., \& {Evans}, N.~J. 2012, \araa, 50, 531, \dodoi{10.1146/annurev-astro-081811-125610}

\bibitem[{{Klein} {et~al.}(2018){Klein}, {Lisenfeld}, \& {Verley}}]{klein_etal_2018}
{Klein}, U., {Lisenfeld}, U., \& {Verley}, S. 2018, \aap, 611, A55, \dodoi{10.1051/0004-6361/201731673}

\bibitem[{{Kulsrud} \& {Anderson}(1992)}]{kulsrud_and_anderson_1992}
{Kulsrud}, R.~M., \& {Anderson}, S.~W. 1992, \apj, 396, 606, \dodoi{10.1086/171743}

\bibitem[{{Lacki} \& {Thompson}(2010)}]{lacki_etal_2010b}
{Lacki}, B.~C., \& {Thompson}, T.~A. 2010, \apj, 717, 196, \dodoi{10.1088/0004-637X/717/1/196}

\bibitem[{{Lacki} {et~al.}(2010){Lacki}, {Thompson}, \& {Quataert}}]{lacki_etal_2010a}
{Lacki}, B.~C., {Thompson}, T.~A., \& {Quataert}, E. 2010, \apj, 717, 1, \dodoi{10.1088/0004-637X/717/1/1}

\bibitem[{{Leroy} {et~al.}(2008){Leroy}, {Walter}, {Brinks}, {Bigiel}, {de Blok}, {Madore}, \& {Thornley}}]{leroy_etal_2008}
{Leroy}, A.~K., {Walter}, F., {Brinks}, E., {et~al.} 2008, \aj, 136, 2782, \dodoi{10.1088/0004-6256/136/6/2782}

\bibitem[{{Longair}(2011)}]{longair_2011}
{Longair}, M.~S. 2011, {High Energy Astrophysics}

\bibitem[{{Manna} \& {Roy}(2023)}]{manna_and_roy_2023}
{Manna}, S., \& {Roy}, S. 2023, \apj, 944, 86, \dodoi{10.3847/1538-4357/acaf64}

\bibitem[{{Moln{\'a}r} {et~al.}(2021){Moln{\'a}r}, {Sargent}, {Leslie}, {Magnelli}, {Schinnerer}, {Zamorani}, {Delhaize}, {Smol{\v{c}}i{\'c}}, {Tisani{\'c}}, \& {Vardoulaki}}]{molnar_etal_2021}
{Moln{\'a}r}, D.~C., {Sargent}, M.~T., {Leslie}, S., {et~al.} 2021, \mnras, 504, 118, \dodoi{10.1093/mnras/stab746}

\bibitem[{{Murphy}(2009)}]{murphy_2009}
{Murphy}, E.~J. 2009, \apj, 706, 482, \dodoi{10.1088/0004-637X/706/1/482}

\bibitem[{{Murphy} {et~al.}(2011){Murphy}, {Condon}, {Schinnerer}, {Kennicutt}, {Calzetti}, {Armus}, {Helou}, {Turner}, {Aniano}, {Beir{\~a}o}, {Bolatto}, {Brandl}, {Croxall}, {Dale}, {Donovan Meyer}, {Draine}, {Engelbracht}, {Hunt}, {Hao}, {Koda}, {Roussel}, {Skibba}, \& {Smith}}]{murphy_etal_2011}
{Murphy}, E.~J., {Condon}, J.~J., {Schinnerer}, E., {et~al.} 2011, \apj, 737, 67, \dodoi{10.1088/0004-637X/737/2/67}

\bibitem[{{Pakmor} {et~al.}(2024){Pakmor}, {Bieri}, {van de Voort}, {Werhahn}, {Fattahi}, {Guillet}, {Pfrommer}, {Springel}, \& {Talbot}}]{pakmor_etal_2024}
{Pakmor}, R., {Bieri}, R., {van de Voort}, F., {et~al.} 2024, \mnras, 528, 2308, \dodoi{10.1093/mnras/stae112}

\bibitem[{{Rieder} \& {Teyssier}(2017)}]{rieder_etal_2017}
{Rieder}, M., \& {Teyssier}, R. 2017, \mnras, 471, 2674, \dodoi{10.1093/mnras/stx1670}

\bibitem[{{Rodrigues} {et~al.}(2019){Rodrigues}, {Chamandy}, {Shukurov}, {Baugh}, \& {Taylor}}]{rodrigues_etal_2019}
{Rodrigues}, L.~F.~S., {Chamandy}, L., {Shukurov}, A., {Baugh}, C.~M., \& {Taylor}, A.~R. 2019, \mnras, 483, 2424, \dodoi{10.1093/mnras/sty3270}

\bibitem[{{Rodrigues} {et~al.}(2015){Rodrigues}, {Shukurov}, {Fletcher}, \& {Baugh}}]{rodrigues_etal_2015}
{Rodrigues}, L.~F.~S., {Shukurov}, A., {Fletcher}, A., \& {Baugh}, C.~M. 2015, \mnras, 450, 3472, \dodoi{10.1093/mnras/stv816}

\bibitem[{{Rubin}(1968)}]{rubin_1968}
{Rubin}, R.~H. 1968, \apj, 154, 391, \dodoi{10.1086/149766}

\bibitem[{{Rybicki} \& {Lightman}(1979)}]{rybicki_and_lightman_1979}
{Rybicki}, G.~B., \& {Lightman}, A.~P. 1979, {Radiative processes in astrophysics}

\bibitem[{{Sargent} {et~al.}(2010){Sargent}, {Schinnerer}, {Murphy}, {Carilli}, {Helou}, {Aussel}, {Le Floc'h}, {Frayer}, {Ilbert}, {Oesch}, {Salvato}, {Smol{\v{c}}i{\'c}}, {Kartaltepe}, \& {Sanders}}]{sargent_etal_2010}
{Sargent}, M.~T., {Schinnerer}, E., {Murphy}, E., {et~al.} 2010, \apjl, 714, L190, \dodoi{10.1088/2041-8205/714/2/L190}

\bibitem[{{Schleicher} \& {Beck}(2013)}]{schleicher_and_beck_2013}
{Schleicher}, D. R.~G., \& {Beck}, R. 2013, \aap, 556, A142, \dodoi{10.1051/0004-6361/201321707}

\bibitem[{{Schober} {et~al.}(2023){Schober}, {Sargent}, {Klessen}, \& {Schleicher}}]{schober_etal_2022}
{Schober}, J., {Sargent}, M.~T., {Klessen}, R.~S., \& {Schleicher}, D.~R.~G. 2023, \aap, 679, A47, \dodoi{10.1051/0004-6361/202245218}

\bibitem[{{Schober} {et~al.}(2013){Schober}, {Schleicher}, \& {Klessen}}]{schober_etal_2013}
{Schober}, J., {Schleicher}, D.~R.~G., \& {Klessen}, R.~S. 2013, \aap, 560, A87, \dodoi{10.1051/0004-6361/201322185}

\bibitem[{{Schober} {et~al.}(2016){Schober}, {Schleicher}, \& {Klessen}}]{schober_etal_2016}
---. 2016, \apj, 827, 109, \dodoi{10.3847/0004-637X/827/2/109}

\bibitem[{{Seymour} {et~al.}(2009){Seymour}, {Huynh}, {Dwelly}, {Symeonidis}, {Hopkins}, {McHardy}, {Page}, \& {Rieke}}]{seymour_etal_2009}
{Seymour}, N., {Huynh}, M., {Dwelly}, T., {et~al.} 2009, \mnras, 398, 1573, \dodoi{10.1111/j.1365-2966.2009.15224.x}

\bibitem[{{Shi} {et~al.}(2018){Shi}, {Yan}, {Armus}, {Gu}, {Helou}, {Qiu}, {Gwyn}, {Stierwalt}, {Fang}, {Chen}, {Zhou}, {Wu}, {Zheng}, {Zhang}, {Gao}, \& {Wang}}]{Shi_etal_2018}
{Shi}, Y., {Yan}, L., {Armus}, L., {et~al.} 2018, \apj, 853, 149, \dodoi{10.3847/1538-4357/aaa3e6}

\bibitem[{{Smith} {et~al.}(2014){Smith}, {Jarvis}, {Hardcastle}, {Vaccari}, {Bourne}, {Dunne}, {Ibar}, {Maddox}, {Prescott}, {Vlahakis}, {Eales}, {Maddox}, {Smith}, {Valiante}, \& {de Zotti}}]{smith_etal_2014}
{Smith}, D.~J.~B., {Jarvis}, M.~J., {Hardcastle}, M.~J., {et~al.} 2014, \mnras, 445, 2232, \dodoi{10.1093/mnras/stu1830}

\bibitem[{{Tabatabaei} {et~al.}(2007){Tabatabaei}, {Beck}, {Krause}, {Berkhuijsen}, {Gehrz}, {Gordon}, {Hinz}, {Humphreys}, {McQuinn}, {Polomski}, {Rieke}, \& {Woodward}}]{tabatabaei_etal_2007}
{Tabatabaei}, F.~S., {Beck}, R., {Krause}, M., {et~al.} 2007, \aap, 466, 509, \dodoi{10.1051/0004-6361:20066731}

\bibitem[{{Tabatabaei} {et~al.}(2013){Tabatabaei}, {Schinnerer}, {Murphy}, {Beck}, {Groves}, {Meidt}, {Krause}, {Rix}, {Sandstrom}, {Crocker}, {Galametz}, {Helou}, {Wilson}, {Kennicutt}, {Calzetti}, {Draine}, {Aniano}, {Dale}, {Dumas}, {Engelbracht}, {Gordon}, {Hinz}, {Kreckel}, {Montiel}, \& {Roussel}}]{tabatabaei_etal_2013}
{Tabatabaei}, F.~S., {Schinnerer}, E., {Murphy}, E.~J., {et~al.} 2013, \aap, 552, A19, \dodoi{10.1051/0004-6361/201220249}

\bibitem[{{Tabatabaei} {et~al.}(2017){Tabatabaei}, {Schinnerer}, {Krause}, {Dumas}, {Meidt}, {Damas-Segovia}, {Beck}, {Murphy}, {Mulcahy}, {Groves}, {Bolatto}, {Dale}, {Galametz}, {Sandstrom}, {Boquien}, {Calzetti}, {Kennicutt}, {Hunt}, {De Looze}, \& {Pellegrini}}]{Tabatabaei_etal_2017}
{Tabatabaei}, F.~S., {Schinnerer}, E., {Krause}, M., {et~al.} 2017, \apj, 836, 185, \dodoi{10.3847/1538-4357/836/2/185}

\bibitem[{{Vollmer} {et~al.}(2022){Vollmer}, {Soida}, \& {Dallant}}]{vollmer_etal_2022}
{Vollmer}, B., {Soida}, M., \& {Dallant}, J. 2022, \aap, 667, A30, \dodoi{10.1051/0004-6361/202142877}

\bibitem[{{Wyder} {et~al.}(2009){Wyder}, {Martin}, {Barlow}, {Foster}, {Friedman}, {Morrissey}, {Neff}, {Neill}, {Schiminovich}, {Seibert}, {Bianchi}, {Donas}, {Heckman}, {Lee}, {Madore}, {Milliard}, {Rich}, {Szalay}, \& {Yi}}]{Wyder_etal_2009}
{Wyder}, T.~K., {Martin}, D.~C., {Barlow}, T.~A., {et~al.} 2009, \apj, 696, 1834, \dodoi{10.1088/0004-637X/696/2/1834}

\bibitem[{{Yun} {et~al.}(2001){Yun}, {Reddy}, \& {Condon}}]{yun_etal_2001}
{Yun}, M.~S., {Reddy}, N.~A., \& {Condon}, J.~J. 2001, \apj, 554, 803, \dodoi{10.1086/323145}

\end{thebibliography}
\bibliographystyle{aasjournal}

\end{document}